*An active hydroelastic liquid crystal phase of a fluttering ferroelectric nematic*

Xi Chen[1], Cory Pecinovsky[2], Eva Korblova[3], Matthew A. Glaser[1], Leo Radzihovsky[1],
Joseph E. Maclennan[1], David M. Walba[3], Noel A. Clark[1]*

[1]*Department of Physics and Soft Materials Research Center,
University of Colorado, Boulder, CO 80309, USA*

[2]*Polaris Electro-Optics, Inc.
3400 Industrial Lane, Suite 7C, Broomfield, CO 80020, USA*

[3]*Department of Chemistry and Soft Materials Research Center,
University of Colorado, Boulder, CO 80309, USA*

*Abstract*

Polarization flutter, produced by an applied AC electric field drives an equilibrium ferroelectric nematic ($N_F$) liquid crystal (LC) through a transition into a dissipative active ferroelectric nematic state exhibiting strong elasto-hydrodynamic intermolecular interaction. In such a fluttering ferroelectric, the typical equilibrium $N_F$ textural features adopted to reduce electrostatic energy, such as preferences for director bend, and alignment of polarization parallel to LC/air interfaces, are overcome, giving way to nonequilibrium conjugate structures in which director splay, and alignment of polarization normal to $N_F$/air interfaces are preferred. Viewing the latter textures as those of an active nematic phase reveals that self-organization to reduce effective viscosity and resulting dissipation generates a flow-driven apparent nematic elasticity and interface structuring that dominates equilibrium LC elastic and surface forces.

*INTRODUCTION*

Soft matter is made "active" by the inclusion of energized elements which are individually driven to move by out-of-equilibrium forces, and which produce macroscopic flow when coupled to fluid degrees of freedom [1,2,3,4]. The major focus of active soft matter theory and experiment has been on biological systems [1], where the driving energy is provided biochemically, with single active elements ranging in scale from the macromolecular [3,5] to the macroscopic [6,7]. Particularly exciting have been the unparalleled opportunities arising for advancing the science of collective active motion [8], wherein soft materials provide flexible systems of interacting active elements that allow facile experimental access, and drive novel theoretical approaches [9].

Here we report a non-biological active soft matter system in which the single active elements are small molecules. We consider the excitation of a thermotropic ferroelectric nematic liquid crystal (LC) comprising RM734, shown in *Fig. 1*, a 2nm long molecular nanorod having an ~11 Debye dipole along its molecular long axes. In equilibrium these molecules organize into the recently discovered ferroelectric nematic phase ($N_F$) [10,11,12,13,14], in which there is nearly-perfect polar order of the dipoles (polar order parameter $p$~ 90%). The $N_F$ is an equilibrium three dimensional (3D) fluid phase having uniaxial nematic order with a substantial macroscopic polarization density (*P*), everywhere parallel to the director $\hat{\bm{n}}$, the local average molecular long axis. Ferroelectric nematic liquid crystals are viscous fluids in which a symmetry breaking phase transition to a state of long-range polar orientational order specified by order parameter and colinear Goldstone variables $\hat{\bm{n}}(\bm{r},t)$ and *P*(*r*,*t*), the three dimensional structure and dynamics of which are determined by internal elastic/hydrodynamic/electrostatic interactions, and applied fields.

In this paper we wed the realms of active soft matter and ferroelectric nematic liquid crystal science. We expose the $N_F$ phase to an oscillating (AC) electric field *E*, such that each molecular dipole *p* experiences an AC torque, $\bm{\tau}_E = \bm{p}\times\bm{E}$, that depends on its instantaneous orientation and neighborhood. The macroscopic effect is a body torque/volume applied the fluid $\bm{\Gamma}_{fl} = n\langle\bm{\tau}_E\rangle = n\bm{p}\times\bm{E} = \bm{P}\times\bm{E}$ where *n* is the molecular number density, and $\langle\bm{\tau}_E\rangle$ the average torque, which induces AC modulation, $\psi(\bm{r},t)$, of the director-polarization ($\hat{\bm{n}}$,*P*) orientation, creating a ferroelectric nematic, with ($\hat{\bm{n}}$,*P*) "fluttering" about average values. This torque-driven molecular reorientation generates flow and, in turn, additional flow-induced torques, a classic scenario that produces the backflow effect [15,16], discovered in the early days of LC display technology development, and successfully explained using LC nemato-hydrodynamic models of Leslie and Ericsson [15,17,18,19,20].

At small $|\psi(\bm{r},t)|$ the induced AC director flutter has little effect on the average director structure of the $N_F$. However, with sufficiently large drive a distinct transition takes place to a new active nematic state in which $\hat{n},P$ self-organizes under a different set of rules. In this state the coupled torque/flow scenario sketched above creates effective elastic and interfacial interaction forces that are orders of magnitude stronger that their equilibrium counterparts, overwhelming them, obliterating the equilibrium $\hat{n},P$ textures, and replacing them with characteristic, essentially different nonequilibrium structural themes. Thus while the equilibrium textures are dominated by bend deformation (to avoid electrostatic self energy), the nonequilibrium textures are dominated by splay. Our analysis shows that in the low-frequency regime, the nonequilibrium textures of the $N_F$ self-organize to reduce the effective viscosity presented to the fluttering drive. This condition for the $N_F$ in-planar cells is achievable only when the insulating layer at the electrode surfaces is capacitive and is the dominant impedance in the cell, which occurs in the low-frequency regime. Such consistent thematic differences that emerge from observations of the simplest textures under AC drive enable the approach to nemato-hydrodynamic analysis employed here, which is to model the dissipation of specific nonequilibrium structural and textural features that drive effective elastic and interfacial forces.

## RESULTS

*Cells* – Experiments were performed using transparent capacitor single pixel sandwich cell geometries, with RM734 at temperature, 60ºC < $T$ < 100ºC, in a $d$ = 0.8 $\mu$m to 2.0 $\mu$m wide gap between glass plates coated with indium-tin oxide electrodes (of area $A$ = 10mm x 5mm), as sketched schematically in *Fig. 1A*. The ITO surfaces (yellow) of the $d$ = 0.8 $\mu$m were bare, with no additional alignment treatment. The LC/ITO structure at the interface creates a $t$ ~ 1nm-thick passive (nonferroelectric) capacitive dielectric layer (lavender) which is insulating and much thinner than the LC layer, and within which $P(r)$ is either absent, or present with fixed orientation. The $d$ = 2.0 $\mu$m cells had $t$ ~ 7nm thick parallel rubbed polymer insulating alignment layers on the plates.

*"Block/polarization"-director ($\hat{n},P$) flutter* – We introduce the electrostatics of such cells, in which a so-called "block polarization"/capacitive Goldstone mode (PCG mode) of molecular reorientation dynamics [21,22,23,24,25] can be driven by applied electric field. The insulating layers in *Fig. 1*, of net capacitance/electrode area $c$, act to separate the polarization charge (blue) at the LC/lavender interface from the free charge (green) at the lavender/electrode interface. If we consider a positive drive voltage, $V_p$ being applied as indicated in *Fig. 1C*, at short times, before there is any motion of *P*, an electric field $E_{LC}(t) = V(t)/d$ appears everywhere in the $N_F$, applying to the uniform *P* a uniform torque/volume, $\Gamma_{fl} = P \times E$. The polarization field then responds by rotating in the direction that transfers charge to the $N_F$ interfaces, of a sign which acts to reduce the field in the $N_F$. If $V(t)$ is less than a saturation voltage $V_{sat}$, then this process ends with the polarization orientation $\psi(V(t))$ such that the interfacial polarization charge $Q(V(t))$ exactly

-3-

cancels (*i.e.*, completely screens) the applied field in the LC, making $E_{LC} = 0$ and, therefore, the torque on ***P*** zero, creating a static equilibrium. The ***P*** field also maintains $\partial P/\partial z = 0$ within the LC layer, eliminating polarization charge and its self-field, making ***P*** and $\psi$ uniform in the LC layer, including overwhelming any surface alignment preference [26], an example of "block polarization switching" [21,22,23,24,25]. For slowly varying fields the quasi-static orientation $\psi(V(t))$ is given by:

$$\sin[\psi(V(t))] = V(t)/(P/c) = V(t)/V_{sat} \qquad (1)$$

where $V_{sat} \equiv P/c = Pd_I/\varepsilon_I$. For $V(t)$ in the range $[-V_{sat} < V(t) < V_{sat}]$ the range of $\psi(V)$ is $[-90° < \psi(V(t)) < 90°]$. For $|V(t)| > |V_{sat}|$ we have saturation, $|\psi(V(t))| = 90°$ and $E_{LC} \propto (V(t) - V_{sat})$, with ***P*** along *z*, the normal to the cell plates. For the cells reported in ***Figs. 2,4-8*** here $V_{sat}$ is in the range $2V < V_{sat} < 6V$, determined from PTOM, as the $V(t)$ where the birefringence $\rightarrow 0$. In ***Figure 3*** for example, $V_{sat} = 3.5V$ for a $d = 0.8$ μm bare ITO cell, $V_{sat} = 4.5V$ for $d = 1$μm bare ITO cell, and in ***Fig. 4*** $V_{sat} = \sim 7V$ for $d = 2$μm cell having 7nm thick polyimide interfacial capacitance layers.

In the dynamic case the uniformity of $\psi(t)$ persists, but, in general, there is a nonzero uniform electric field, $E_{LC}(t)$ in the LC, so that $\dot{\psi} = PE_{LC}(t)/\gamma$ [13,14], and $\gamma$ is the effective nematic orientational viscosity for director reorientation ($\gamma$ equals the LE $\gamma_1$ in absence of flow of the LC, but will be smaller if the reorientation drives flow) [20]. If $V(t)$ is sinusoidal $[V(t) = (V_p)_\omega e^{i\omega t}]$ then, for small $\psi$, the dynamic version of ***Eq. 1*** is [24]:

$$(E_{LC})_\omega = [i\omega\tau_o/(1 + i\omega\tau_o)] (V_p)_\omega/d, \qquad (2a)$$

$$\dot{\psi}_\omega = i\omega\psi_\omega = [P/\gamma_{eff}](E_{LC})_\omega = (\Gamma_{fl})_\omega/\gamma_{eff}, \qquad (2b)$$

$$\psi_\omega = [(V_p)_\omega/V_{sat}][1 + i\omega\tau_o]^{-1}, \qquad (2c)$$

$$\dot{\psi}_\omega \approx i\omega[(V_p)_\omega/V_{sat}] \qquad \text{for } (\omega\tau_o < 1), \qquad (2d)$$

where $(\Gamma_E)_\omega = P(E_{LC})_\omega$ is the applied electric torque density, $V_p$ is the peak amplitude of $V(t)$, $\psi_\omega$ is also independent of *z*, $\gamma_{eff}$ is the effective orientational viscosity of the $N_F$, $\tau_o = R_{eff}C$, $C = cA$, where $A$ is the area of the cell electrode, and $R_{eff} = \rho_{eff}d/A$ is the effective resistance of the $N_F$ layer, with $\rho_{eff} = 1/\sigma_{eff} = \gamma_{eff}/P^2$ [21,24]. For RM734 at $T = 100°C$, high-speed switching measurements where the fluid does not flow gives $P = 6 \times 10^{-2}$ C/m$^2$ and $\gamma_{eff} = \gamma_1 = 0.5$ Pa-s [13]. Taking $d = 0.8$ μm, $A = 0.5$ cm$^2$, we find $R_1 = \rho_1 d/A = [\gamma_1/P^2]d/A = 25$ Ω, in agreement with our measured value of 22 Ω, *i.e.*, we find $\gamma_{eff} \approx \gamma_1$. Thus the bulk electrical impedance of the $N_F$ layer is polarization resistive, a result of the viscous overdamping of the reorientation of ***P***, and the relaxation time $\tau_o$, which also depends on $\gamma_{eff}$, is an "RC" time constant of the effective cell impedance: the cell surface capacitance $C = cA$ (for the $d = 0.8$ μm cell we estimate $C \sim 1\mu F$ with $\varepsilon_I \sim 3$, $d_I \sim 1$nm, $A = 0.5$ cm$^2$) in series electrically with the cell bulk resistance $R = 22$ Ω. These equations apply identically to the free-boundary single block, and sticky-boundary double and triple block geometries of ***Fig. 1F-H***.

The frequency $\omega = \tau_o^{-1} \sim 5\text{x}10^4$ Hz marks the crossover from an interfacial $C$-dominated cell impedance at low frequency, to an $R$-dominated cell impedance at high frequency. Resistance $R$ means that in the N$_F$ phase the LC is an electrical conductor in the $z$ direction, its conductivity, due to overdamped polarization reorientation, given by $\sigma_1 = P^2/\gamma_1$. This conductive medium shares the known properties of electrical conductors [27] including, for example, that the N$_F$ volume will relax to become charge-free in equilibrium, with $E_z = 0$ everywhere inside. Active N$_F$ behavior is a phenomenon of the in the low-frequency, $C$-dominated regime. Active N$_F$ behavior is a phenomenon of the low-frequency, $C$-dominated regime, where the induced reorientation depends only on the applied voltage, and spontaneous textural reorganization is directed toward reducing viscous drag and dissipation.

***d = 0.8 μm random planar cell*** – The as-prepared bare ITO cell surfaces orient the $\hat{n}$,**P** couple of the N$_F$ to be generally parallel to the surface but uniform in local orientation only over small few-μm length scale rice-grain-shaped domains in the ($x,y$) plane. Without drive voltage we have $\psi(V(t)) = 0$ (***Fig. 1B***) and $\hat{n}$,**P** is held in the bulk LC parallel to the cell plates by the electrostatic energy $U_e = \frac{1}{2}(PV_{sat}/d)\psi^2$ [24], enforcing the random planar alignment induced by the surface, deforming the elastic nematic ground state, to generate a typical (Schlieren) texture, imaged in ***Figs. 2A,D*** using depolarized transmission optical microscopy (DTOM). This texture is locally uniaxially birefringent, giving transmission of depolarized incident light in the lowest order grey/white band of a Michel-Levy chart [28], as expected for RM734 ($\Delta n \sim 0.2$), in a $d = 0.8$ μm thick cell.

***Fluttering Ferroelectric Smooth texture (FFS texture) at T = 100 ºC*** – Upon introducing $\hat{n}$,**P** flutter into these cells, with sinusoidal drive amplitude ($V_P = 10$V, $\omega = 900$ Hz) and $T = 100$ ºC, a striking transformation in the LC texture takes place, as shown in ***Fig. 2***. The cells transition from the random planar disorder (***Fig. 2A***), to a very smoothly varying birefringent texture, a change which is reversible upon removal of the field (***Fig. 2D***). ***Fig. 2B*** shows an intermediate cell condition, 0.016 sec after field application, in which a dynamic phase front is moving upward on the image leaving behind the Fluttering Ferroelectric Smooth texture (FFS texture), a process taking ~0.03 sec to complete. The birefringent contrast of the disordered random planar texture is eliminated where this FFS texture pattern has appeared, indicating that the internal effective elastic forces stabilizing the FFS texture dominate the surface alignment interactions, filling the cell from surface to surface with the FFS texture pattern. Upon removal of the field the random planar domains return in ~0.03 sec to an equilibrium state very similar to that before field application, except that now the local orientation of the rice-grain domains is somewhat biased to be along what was the local FFS orientation. This is a form of surface memory effect [29], which is observed in equilibrium systems when large in-plane torques imprint a pattern of preferred azimuthal orientations on a LC/solid interface, apparently also the case here.

The FFS textures of *Figs. 2,4,5* are found at T = 100 °C for $f$ in the range (20 Hz < $f$ < 2kHz) and $V_p$ ≳ $V_{sat}$, that is in the low-frequency regime with driving such that $\psi(V(t))$ approaches 90° through each drive cycle. For $f$ > 2kHz another type of domains appear which are not the focus of this paper.

The FFS texture is further evident in *Fig. 4*, obtained in a $d$ = 2μm cell with thicker (7nm) rubbed polymer alignment layers giving the untwisted monodomain in *Fig. 4A* with $V_d$ = 0 . With $f$ = 1000 Hz, $V_p$ = 10 V drive, the FFS texture obtained reveals no observable evidence for this uniform surface condition, as can be seen in *Fig. 4B*. Additionally, upon removing the drive, it can be seen that the surface alignment conditions of *Fig. 3A* have been substantially and permanently modified by their contact with the fluttering nematic, such that the majority of the cell area remains in spatially homogeneous states where the $\hat{n}(r),P(r)$ field is tilted at a fixed angle away from the $z$ axis, forming heliconical, lower birefringence π twisted states of opposite left (LH) and right (RH) handedness, separated by faceted domain walls. When heated up back to the N phase, the twist is lost but areas remain with the director tilt maintaining a lower birefringence compared to the initial planar monodomain. The FFS state of *Fig. 4B* returns with field application from any of these field-free conditions.

<u>*FFS local $\hat{n},P$ structure*</u> – Once the cell was filled with FFS textures as in *Figs. 2,4* the smooth FFS areas were evaluated using DTOM with a variable compensator, while local polarized laser illumination directly probed the local $\psi(V(t))$. Reorientation of these cells between crossed, and symmetrically uncrossed polarizer and analyzer shows that $\varphi(x,y)$, the azimuthal orientation of $\hat{n}$ $(r),P(r)$ is independent of $z$, showing in *Fig. 2I* no evidence for twist through the cell.

The FFS texture was probed in detail in the $d$ = 0.8 μm cell, and indeed found to be a fluttering one, as quantified in *Fig. 3A* which shows a direct probe of the flutter, using the transmitted intensity, $I(t)$, of a laser beam, focused to the green ring location in *Fig. 2C*, polarized at 45° from the local $\hat{n},P$ orientation, and passing through the cell at normal incidence, plotted as a function of the instantaneous drive voltage, *V*(*t*). The resulting ψ(*V*(*t*)) was probed by measuring sample birefringence which gives the "V-shaped" optical transmission curve between crossed polarizer and analyzer expected for block polarization reorientation [25]. These data yield $|V_{sat}| = 3.5$ V. The resulting red curve giving ψ(*V*(*t*)) agrees with *Eq. 1*, for $V < V_{sat}$, except for a small voltage shift which returns ψ(t) to zero at $V(t) \neq 0$, due to resistive leakage of the interface capacitors.

In order to probe the instantaneous tilt of the local optic axis out of the cell plane during fluttering, this this laser experiment was also carried out with the cell tilted through ~25° about an axis parallel to the laser polarization, for four different azimuthal directions of ***n̂***(*r*),***P***(*r*) in the cell plane, obtained by moving the laser to the different positions shown in *Fig 2J* (black, green, blue, red) in the radial ***n̂***(*r*),***P***(*r*) field of a +1 defect. Here the dashed yellow line is the cell tilt axis, so that if ***n̂***,***P*** is locally tilted (say ***P*** is tilted out of the cell plane toward the reader) then the DTOM image in *Fig 3B* would show three distinct birefringence values (black=blue, green, red). However, if ***n̂***,***P*** is parallel to the cell plane then there will be only two (black=blue, green=red), as is quantified in *Fig. 3B*, showing that the apparent optic axis *remains parallel to the cell plane during all phases of fluttering*, a result not consistent with the single block polarization switching geometry sketched in *Fig. 1B*, where the optic axis must tilt out of the cell plane, but rather indicating that the $N_F$ structure is double or triple block ("double" as in Fig. 1D), in which the local instantaneous tilt of ***n̂***,***P*** averages to give an untilted apparent optic axis.

*FFS induced flow* – This optical result leads to consideration of the local fluid flow patterns induced by fluttering. As discussed in deriving *Eqs. 1,2*, during block polarization with φ(*x,y*) uniform, the fluttering ***P***(z) is spatially uniform to eliminate space charge ∂***P***/∂z. The AC field in the liquid crystal, $(E_{LC})_\omega$, applies uniform AC body torque density $(\Gamma_E)_\omega = P(E_{LC})_\omega$ to the ***n̂***,***P*** field, which in turn appears as body torque on the fluid, $(\Gamma_{fl})_\omega = (\Gamma_E)_\omega$, driving fluid flow velocity ***v***(*r*). For low Reynolds number and steady flow of an incompressible fluid the Navier-Stokes equation gives $\gamma_{eff} \nabla \times \nabla \times v(r) + \nabla \sigma = \nabla \times \Gamma_{fl}$, where **σ** is the stress tensor and $\nabla \times \Gamma_{fl}$ is the effective body force due to the applied torque [30,31]. In the one dimensional case of *Fig. 1,* where the velocity is $v_u(z)$, we have $\nabla \times v(r) = t[\gamma_{eff}(\partial v_u(z)/\partial z) + \sigma_u] = \Gamma_{fl}$ as sketched in *Fig. 1E*: the uniform deposition of torque density drives either a linear velocity gradient or a shear gradient, depending on the boundary conditions. If, in a cell, $v_u(z)$ has slip boundary conditions, i.e. $\sigma_{uz} = 0$ at the plate, then a single-block reorientation, as in *Figs. 1B* and *F* with a linear velocity gradient $\gamma_{eff}(\partial v_u(z)/\partial z) = \Gamma_{fl}$ and stress $\sigma_{uz} = 0$ fills the cell, the minimum dissipation state. However, if, as is typically the case, the fluid sticks at the surfaces ($v_u(z) = 0$), then in single block reorientation $\partial v_u(z)/\partial z = 0$ and $\sigma_u = \Gamma_{fl}/\gamma_{eff}$, the maximum dissipation state. Dissipation can be reduced to the former level if either the two- or three-block reorientation modes, (*Fig.1 D,G* or *H*, respectively) are driven instead. In these cases the block switching will have the following features: (*i*) $\hat{n}(z)$, $P(z)$ and $v_u(z)$ are confined to the local ($u$,$z$) plane, which is normal to the plates; (*ii*) $v_u(z) = 0$ boundary condition at the electrode plates In the two and three block flows this is equivalent to effectively free boundaries on each block; (*iii*) $\sigma_{uz} = 0$ everywhere; (*iv*) $|\partial v_u(z)/\partial z|$ = $|\Gamma_{fl}/\gamma_{eff}|$; (*v*) alternating signs of $\psi$, $\Gamma_{fl}$, and $|(\partial v_u(z)/\partial z)|$ in adjacent blocks; (*vi*) polarization stabilized kink walls [26] between blocks, which walls disappear at $\psi = 90°$; (*vii*) continuity of velocity $v_u(z)$ at the block walls; (*viii*) direct applicability of *Eqs. 1,2* describing the electrostatics of block polarization to both single- and multiple- block switching; (*ix*) flow stabilization of sharp interfaces between blocks; (*x*) zero net charge separation along *u* across the cell plane in the multi-block cases, *i.e. Fig. 1D* rather than *Fig. 1B*; (*ix*) universal applicability of *Eqs. 3-5* to describe ferroelectric nematodynamics in single- and multiple- block switching. Finally, we note that two-block switching generates net transverse net fluid flux (*Fig. 1G*), while in (*Fig. 1H*), flow in the upper half of the cell, is cancelled by that in the lower half, accommodating zero net flux (*Fig 1H*), the case expected if net flow is blocked, for example if $\hat{n}$,$P$ is normal to a boundary. We can estimate the resulting net displacement of fluid at low frequency in the two block case by using *Eqs. 2d,4*, which, for $V_p \sim V_{sat}$, have $\dot{\psi}_\omega = \partial v(z)/\partial z \approx i\omega(H)$, where $H < 1$ is a dimensionless ratio of viscosities, given below. This estimates the peak-to-peak displacement $\delta u$ of fluid by AC flow along *u* in any of the geometries of *Fig. 1* to be $\delta u \lesssim d$.

*Broken azimuthal symmetry* - It is important to point out that while the electrostatic arguments leading to *Eq. 1* constrain both $P_z$ and $\partial P_z/\partial z$ {$P_z = cV(t)$, $\partial P_z/\partial z = 0$}, the spontaneously broken symmetry in azimuthal orientation of $\hat{n}$,$P$ in the ($x$,$y$) plane persists in the FFS, with its Goldstone variable $\varphi(x,y,z)$ determined solely by internal elastic and dissipative interactions. In both these cell types experiments show that the FFS texture is characterized by: (*i*) a $\varphi(x,y)$ that is independent of $z$, showing little evidence for twist through the cell in the FFS textures of *Figs. 2,4-8*), giving the preferred local state of $z$-independent $\varphi(x,y)$ sketched in *Figs. 1G,H*. (*ii*) a continuous variation of $\varphi(x,y)$, the azimuthal orientation of the ($u$,$z$) plane of $\hat{n}$,$P$ flutter (*Fig. 1*), giving the projection of the local orientation of $\hat{n}$,$P$ in the ($x$,$y$) cell plane. (*iii*) A minimum population of $+2\pi$ topological or other defects, keeping only those required by the boundaries, all features pointing to stabilization of the FFS texture by effective orientational elasticity and interfacial interaction that is dominant.

*Unexpected features of the FFS texture* – The maps of $\hat{n}(r)$,$P(r)$ in FFS textures in *Figs. 2C,4B* show that the most commonly occurring in-plane *n*,*P* structures of the FFS texture are $+2\pi$ topological singularities, with the polarization *radial*, pointing either in toward the center or out, and *splayed*. Additionally, the orientation of $\hat{n}$,$P$ varies continuously such that $P(r)$ tends to terminate everywhere *normal* to the boundary lines of the LC with the black areas which are bubbles between the plates. This self-organization of $\hat{n}(r),P(r)$ is surprising because: (*i*) Equilibrium textures obtained in absence of boundary torques on planar bounding surfaces, for example with free boundaries on fluid support [32] or in freely suspended films [33], exhibit +2π defects with no tendency to be radial and splayed, but are rather *tangential* and *bent*. This is because splay of *P(r)* generates polarization charge $\rho_P(r) = -\nabla \cdot P(r)$, while bend of *P(r)* has $\nabla \cdot P(r) = 0$, thus avoiding space charge [13,27,34,35,36]. (*ii*) The preferred equilibrium orientation of $\hat{n}(r),P(r)$ at a dielectric boundary is therefore tangent to the boundary [13,27,34]. In virtually every publication where equilibrium $N_F$ textures are shown and polarization direction analyzed, *P(r)* is found to be *parallel* to LC/air interfaces, a simple geometrical preference that reduces the electric field in the air, and thus the electrostatic energy, in a fashion analogous to the triangular magnetic surface domains in ferromagnets [37,38]. Geometries where *P(r)* is aligned parallel to LC/air surfaces include $N_F$ freely suspended filaments [39,40,41,33] and films [33], free drops on surfaces [42], and, most relevant here, numerous direct observations of bubbles in cells, where the in-plane orientation of $\hat{n}(r),P(r)$ is tangent to the bubble boundary (***Fig. 4A***).

*Periodic pattern formation in FFS textures* – Extended defect-free lattices of the ***Fig. 2*** FFS textural motif can be readily induced at $T = 100$ °C for *f* in the range (20 Hz < *f* < 2kHz) and $V_p > 5$V. ***Fig. 2*** shows that in the FFS texture/bubble geometries the principal spontaneous self-organizational theme is to fill patches of open areas of dimension ~200 to 300 μm with a +2π radial splay defect of finite size, bounded by bubble interface lines (magenta shading) or splay/bend defect lines (cyan, discussed below). A possible interpretation of this observation is that there is a preferred range of splay curvature magnitude, $S_{max} > S > S_{min}$, and, since $S \sim 1/\rho$ where $\rho$ is the distance from the core, the outer reaches of the defect become unfavorable. Thus, over larger areas the system prefers arrangement of elemental 2π defect cells of finite size. These behave like particles in the 2D plane of the LC layer which mutually attract with adhesive cyan boundaries holding them together, typified by that in ***Fig. 2C,E,F,G***, where they organize around random bubbles. ***Fig. 2G*** shows a transient example of four +2π radial defects associating around the crossing point of a pair of crossed cyan defect lines, an assembly which creates a topologically compensating -2π defect. This motif, stabilized here by the bubble boundaries in dynamic local arrangements, is much like the dynamic association of multifunctional particles which form colloidal crystals [43], and indeed, at drive higher than that of ***Fig. 2***, the metastable arrangement of ***Fig 2G*** is established as the unit cell of 2D periodic arrays of the basic +2π defect, represented by the colored plaquette, as shown in ***Fig. 5E***, as part of an extended lattice, obtained with a $V_p = 5$V triangle wave at *f* = 900 Hz. The weak remnant birefringence at the largest voltages (***Fig. 5E***) depends on the sign of the field, showing that nearest neighbor +1 defects have alternate polarization sign.

Periodic director textures very similar to that in *Fig. 5* have been previously observed in nematic cells as a Rayleigh-Benard type of instability when the nematic is heated from above [44,45]. This thermal instability is successfully described theoretically by spatially periodic solutions of the coupled orientation-flow nematohydrodynamic equations including temperature variation [44,45]. Similar periodic lattice textures have also been found in nematics having negative dielectric anisotropy and subjected to an AC electric field in sandwich electrode cells [46,4748,49]. This electrohydrodynamic instability has not been treated theoretically, although periodic lattice solutions to a nematic hydrodynamic model have been demonstrated [50].

*FFS and equilibrium textures are conjugate* – We note that the spontaneously adopted orientation distribution of $\hat{n},P$ in the topological defects and near the boundaries of the FFS texture, apparently uninfluenced by surface forces, is essentially conjugate to that which an equilibrium $N_F$ with similarly free surfaces and under the influence of its internal elastic and electrostatic interactions would adopt. Thus, carrying out a local 90º reorientation of $\hat{n},P$ everywhere in the textures of *Figs. 2,4B,5* generates an equilibrium-like texture from a driven FFS texture [32,33]. In equilibrium the dominant electrostatic space charge energy cost suppresses $\nabla \cdot \hat{p}(r)$, suggesting that $\nabla \times \hat{p}(r)$ is suppressed in the FFS.

*Fluttering Ferroelectric Smooth texture (FFS texture) at T = 43ºC* – At lower temperatures the RM734 viscosity is much larger [13,14], enabling the generation of fluttering stresses at much lower strain rates. As a result, FFS textures can be stabilized in the $d = 0.8$ μm bare ITO cells at much lower frequencies, with *Figs. 6-8* showing areas of periodic FFS textures obtained at $T = 43$ ºC and frequencies $f = 0.2, 1.0,$ and $4.0$ Hz. These exhibit the strong dependence of lattice period $L(f)$ on frequency, shown in *Fig. 7A*. Additionally, the surface memory effect [29] becomes stronger at lower temperatures, such that under driving conditions it can dynamically apply stresses comparable to those of the flow, enabling the surface to template particular FFS patterns, including periodic lattices, while they are being established by drive at a particular frequency and making it possible to follow the dynamics of the change within fluttering cycles as in *Figs. 6,8*, and between them as the frequency is changed, as shown in *Fig. 7B*. However, the influence of the flow on the surface is largely lost upon removal of drive even at low *T*, as the random planar texture reappears, with a weak bias remembering the prior local flow direction, as seen in the transition from *Fig. 8A* to *8B*.

Experimentally, these observations show that transition to the FFS phase takes place for $V_p > V_{sat}$, meaning that, in the low frequency regime, the fluttering cycle of the FFS employs the full angular range of $\psi$ (-90º < $\psi$ < 90º), as in the high frequency regime, shown by the birefringence measurements in *Fig. 3*. The images in *Figs. 2,4* are, at high *f*, time-averaged textures, recorded at drive frequencies much larger than the 60 Hz video frame rate. By contrast, the low-$T$ experiments at the lower frequencies also make possible the experiments in *Figs. 6-8*, in which FFS textures can be probed at various times through single fluttering cycles.

Observations at $T = 43$ °C can be summarized as follows:

(*i*) <u>pattern variation during AC flutter strokes</u> – The typical time variation within each cycle of such a stabilized FFS array, established by a periodic-in-time fluttering drive, is shown in *Fig. 6* in a sequence of DTOM images obtained over a 5 sec-long sample (~1 fluttering cycle at $f = 0.2$ Hz, with $V_p > V_{sat}$). Each row is taken at equal intervals, with some black images deleted at the extreme $|\psi| = 90°$ orientations. These images show directly the dynamic stabilization of the patterns, as they become noticeably more diffuse for large $\psi$ but are become more sharply defined by the return of $\psi$ to $\psi = 0$ in the cycle.

(*ii*) <u>frequency dependence of the lattice plaquette size, L</u> – DTOM images of typical FFS periodic array textures obtained at different frequencies are presented in *Fig. 7A*, where each image is captured at $\psi = 0$, the brightest place in a cycle. We find that the FFS textures are dynamically stabilized by fluttering at each $f$-value, with $L$ showing a monotonic increase of the $+2\pi$ defect plaquette size with decreasing frequency, varying approximately as $L \propto \sqrt{(1/f)}$.

(*iii*) <u>pattern variation in response to step change of frequency</u> – This sequence of images in *Fig. 7B* follows the evolution of an initial pattern driven at $f = 1$ Hz having a $+2\pi$ defect plaquette dimension $L \sim 30$ μm, to a new pattern stabilized with $L \sim 60$ μm, following a step change in frequency at $t = 0$ to $f = 0.2$ Hz. The notable changes from cycle to cycle do not take place at times when $\psi$ is small and the image bright, but rather for $V \gtrsim V_{sat}$ where the image is dark.

(*iv*) <u>active patterning from disorder within a single flutter stroke</u> – At $T = 43°$ the periodic pattern of *Fig. 8A* is obtained with drive $f = 0.5$ Hz, $V_p = 10$V. However if $V_p$ is switched to zero, then over the few minute long interval following, the periodic pattern becomes diffuse, nearly disappears, and is replaced by the random planar speckle texture of *Fig. 8B*. If $V(t)$ switched back on, the original periodic pattern reappears, stimulating a study of the dynamics of this process. We observe that, in the time interval following this switch-on at $t = 0$ the passage through a single + or - peak of $V(t)$ [*Fig. 8B→C*], and then back to $V(t) = 0$ [*Fig. 8C→D*], is sufficient to completely eliminate the random domains and rewrite the long-term stabilized FFS texture of *Fig. 8A* almost perfectly. Continuing through the next half cycle changes the pattern very little, only making the DSK lines in the texture slightly sharper. Thus the pattern stabilized at $f = 2$ Hz in *Fig. 8*, effectively being rewritten each half cycle ending with $V(t) = 0$, is that for which dynamically traversing the $\psi(t)$ trajectory from $\psi = 90°$ to $0°$ in exactly 0.5 sec applies no torque $\Gamma_z$ tending to change the pattern $\varphi(\mathbf{r},t)$. Slower or faster traverses will apply such torques, tending to expand or contract the lattice cell size, respectively.

*Stabilization of the FFS state by flutter* – We propose that the FFS texture is a nonequilibrium (active nematic) state created and stabilized by $\hat{n},P$ flutter. The possibility that such an FFS state could exist can be appreciated by considering the net Frank elastic energy of a typical nematic texture in the context of the power being fed into the nematic flow by flutter. The former can be estimated for the Schlieren texture of *Fig. 2A* as $U_K = [\frac{1}{2}K(\pi/d)^2|\psi|^2]Ad$ for a $d \sim 1\mu m$, $A = 0.5 cm^2$, Frank constant $K \sim 5 \times 10^{-12}$ N cell, with $|\psi|^2 \sim 1$. This estimate gives $U_K \sim 10^{-8}$ Joules for the LC texture in the cell. The power being deposited into hydrodynamic flow of the $N_F$ by fluttering is just the electrical power $W_{fl} = \frac{1}{2}\text{Re}\{[(V_p)_\omega/Z_\omega](V_p)_\omega^*\} = [\frac{1}{2}V_{sat}^2/R][(\omega\tau_o)^2/(1 + (\omega\tau_o)^2)]$, flowing into the effective damping resistance $R$ of the $N_F$. Here $Z_\omega = (R + 1/i\omega C)$, and, from above, $R = \rho d/A = [\gamma_1/P^2]d/A = 22 \Omega$. With $V_{sat} \sim 5V$ and $\omega\tau_o \sim 1$, and $V_p \sim V_{sat}$ so $|\psi|^2 \sim 1$, we have $W_{fl} \sim 1$ Watt, putting in one equivalent of the cell's elastic energy every 10ns. This large energy flux is made possible by ferroelectricity, the polarization of the $N_F$ enabling large driving torques at moderate applied voltages as shown by comparison of typical elastic torque per unit area in the texture, $K(\pi/d) \sim 3 \times 10^{-5} J/m^2$, with the field-applied value, $PV_{sat} \sim 0.3 J/m^2$. These torques are equally active on both signs of the AC driving field, a key advantage over dielectric nematics in electrohydrodynamic driving.

*Flow-alignment by flutter & the local structure of the FFS texture* – Comparison of equilibrium textures with nonequilibrium FFS textures, for example in *Figs. 1A* and *C*, *Figs. 4A* and *B*, and in *Fig. 8B* and *D* show that the nonequilibrium stability, spatial variation, and dynamics of $\varphi(x,y)$ are largely self-determined by internal interactions of $\hat{n},P,v_u$, and $\Gamma_{fl}$, where we take the in-plane electric field to be zero. Based on the discussion above, we now consider the local fluttering dynamics and their manifestation in the structure of $\varphi(x,y)$ in the FFS textures, starting with an untwisted texture of uniform $\varphi(x,y)$ executing multi-block fluttering in the $(x,z)$ plane, with stick ($v_u = 0$) boundary conditions as in *Figs. 1G,H*. According to *Figs. 4,6,8* some observed phenomena have $|\psi|$ approaching 90°, and so are clearly in the nonlinear regime of *Eq. 1*, indicating that a full treatment of this problem requires numerical or simulation solution of the nonlinear nematic electrohydrodynamics equations [15,16]. The treatment pursued here is limited to small $\psi$, where linear analytic expressions for fields and the resulting dissipation are readily obtained, and can be used to understand the apparent nonequilibrium elastic behavior of $\varphi(x,y)$.

Within each of the two or three blocks, small amplitude sinusoidal flutter with $\varphi(\mathbf{r}) = 0$ then applies to $\hat{n}(\mathbf{r}),P(\mathbf{r})$ a $z$-independent sinusoidal torque density field $(\Gamma_{fl})_\omega = \hat{p} \times \hat{z}(\Gamma_{fl})_\omega$, of magnitude $(\Gamma_{fl})_\omega = P(E_{LC})_\omega$, normal to $\hat{n}(\mathbf{r}),P(\mathbf{r})$ and having frequency dependence given by $(E_{LC})_\omega$ in *Eq. 2*:

$$(\Gamma_{\text{fl}})_\omega = P(E_{\text{LC}})_\omega = [P(V_{\text{p}})_\omega/d][i\omega\tau_{\text{o}}/(1 + i\omega\tau_{\text{o}})] = [P(V_{\text{p}})_\omega/d][R/(R + Z_{\text{C}})], \quad (3a)$$

$$D_{\text{fl}} = \tfrac{1}{2}\text{Re}\{[(\Gamma_{\text{fl}})_\omega \dot{\psi}_\omega^*\} = [\tfrac{1}{2}|V_{\text{p}}|^2/R][(\omega RC)^2/(1 + (\omega RC)^2)]/Ad \quad (3b)$$

where $Z_C = 1/i\omega C = 1/i\omega cA$, $R = [\gamma_{\text{eff}}/P^2]d/A$, where $\gamma_{\text{eff}}$ is the effective viscosity opposing the torque, and $D_{\text{fl}}$ is the resulting dissipated power flux/volume. Of importance to note here is that at low frequency ($\omega\tau_{\text{o}} < 1$) we have $D_{\text{fl}} \propto R \propto \gamma$, so that reducing effective viscosity means reducing the dissipation, apparently the basic condition that stabilizes the FFS. At high frequency ($\omega\tau_{\text{o}} > 1$), since $D_{\text{fl}} \propto 1/R \propto 1/\gamma$, the opposite is true.

The standard nemato-hydrodynamic equations, summarized in Supplemental Information *Section S1*, couple $\dot{\psi}$ and $\Gamma_{\text{fl}}$ and to the shear stress in the fluid, $\sigma(z)$ and velocity field $v(z)$. The $z$-independence of $\dot{\psi}$ and $\Gamma_{\text{fl}}$ make $\sigma$ also independent of $z$, and, taking $\hat{u}$ along $\hat{x}$ make $v_{\text{u}}(z) = v_{\text{u}}(z)\hat{x}$ vary linearly with $z$, defining $G_{\text{fl}} \equiv \partial v(z)/\partial z$ as the resulting $z$-independent shear velocity gradient in the LC. Assuming *planar* alignment as in *Fig. 1E*, small magnitude of $\psi$ yields coupled linear equations giving $\dot{\psi}$ and $G_{fl}^{planar}$ in terms of $\Gamma_{\text{fl}}$ and $\sigma$. In multi-block fluttering $v_{\text{u}}(z)$ experiences free-slip boundary conditions for at each of the internal interfaces between blocks in *Fig. 1G,H*, which is introduced simply by setting $\sigma = 0$. In this case within each block the velocity varies linearly with $z$, and the block interfaces will position themselves along $z$ such that $v_{\text{u}}(z) = 0$ at the pink/grey electrode surfaces (at $z = d/2$ for two blocks, and at $z = d/4$, $3d/4$ for three blocks). In each block the magnitude of the gradient of $v_{\text{u}}$ will be $|(\partial v_{\text{u}}(z)/\partial z)| = (G_{fl}^{planar})$, where (Supplementary Information, [20,51,19]:

$$(G_{fl}^{planar})_\omega = [P(E_{\text{LC}})_\omega/\gamma_{\text{HF}}][\alpha_3/\eta_2] = i\omega[P/\gamma_{\text{HF}}][\alpha_3/\eta_2][i\omega\tau_{\text{oHF}}/(1 + i\omega\tau_{\text{oHF}})](V_{\text{p}})_\omega/d \quad (4a)$$

$$(G_{fl}^{HF})_\omega \approx i\omega[\alpha_3/\eta_2][(V_{\text{p}})_\omega/V_{\text{sat}}] = [\alpha_3/\eta_2][I_\omega/P] = [\alpha_3/\eta_2][\dot{\psi}_\omega] \quad \text{for } (\omega\tau_{\text{oHF}} < 1). \quad (4b)$$

Here $I_\omega$ is the cell current; HF indicates "hard flow", defined in the next section; and $\tau_{\text{oS}} = R_{\text{S}}C$ is geometry-dependent. In this planar case $R_{\text{HF}} = [\gamma_{\text{HF}}/P^2]d/A$, where $\gamma_{\text{HF}} = \gamma_1[1 - (\alpha_3/\eta_2)^2]$, $\approx \gamma_1$, since $\alpha_3$ is typically small in magnitude. The effective viscosity dissipating the input power in *Eq. 3B* is $\gamma_{\text{eff}} = \gamma_{\text{HF}}$, smaller than $\gamma_1$ because the extent to which the fluid can simply rotate in response to the torque, represented in nonzero $G_{\text{fl}}$, reduces $\gamma$.

*"Hard Flows" (HFs) and "Easy Flows" (EFs)* – An emergent general theme is that the AC-generated driven flows of the fluttering state can produce time-average DC (quasi-static) forces and torques which depend on, and can alter, the local structure, and spatial variation of $\varphi(r)$. We now describe such behavior in the fluttering $N_F$, using the notions of what we call "hard flows" (HFs) and "easy flows" (EFs), beginning with the thought experiment shown in *Fig. 6A*. Here a floating motorboat arranged to be normal to, and to have its bow in contact with, a smooth rigid wall. Its motor, which can be viewed as an underwater flipper-like paddle is running, providing maximum motive power but there is little velocity: a "hard flow (HF)", high dissipation situation, in fact the hardest HF limit. If the boat is not quite normal to the wall it will experience on average nonzero (DC) reactive force from the wall that tends to gradually turn it to be sliding along the wall, beginning a reorientation event that leaves the boat nearly parallel to the wall. Now, the motor can, with much less dissipated power, move the boat along the wall at a comfortable speed, having achieved a state of easier flow, eventually reaching that in open water, which would be the lowest dissipation situation, the "easy flow" (EF) limit. The system spontaneously finds and evolves to the easiest flow situation that is accessible, the elastic -like turning forces coming from the difference in the net force exerted on the boat by the wall between the forward and return strokes of the paddle.

*Fig. 6B,C* shows two basic HF and EF geometries of fluttering ferroelectric nematics having uniform $\varphi(r)$, and having the resting $\hat{n},P$ either parallel or normal to the plates, respectively planar or homeotropic, as indicated. Consider this LC layer to be the central block in *Fig. 1H*, where it has effectively free boundary conditions on $v_u$, and $\sigma = 0$. In such a volume of uniform director orientation, the flow velocity is a simple linear shear that must be uniform in direction and amplitude, and, because of the translational symmetry along $\hat{u}$, must be parallel to the plates, as sketched in *Fig. 1G,H*. A driven molecule serves as oscillating low Reynolds number paddle, which most efficiently couples to fluid velocity fields normal to the surface of its extended length. Of the two geometries in *Fig. 6B,C* only the homeotropic one matches the direction of this generated velocity with that of the permitted simple shear flow velocity, making this geometry the limit of easy flow (EF limit). By contrast, in the experiments here, where $\hat{n},P$ are *planar* aligned, the efficiently generated flow would be normal to the plates, but flow along $\hat{z}$ is obviously blocked by the plates, and by the translational symmetry along $\hat{u}$ under the continuum condition of *Fig. 1G,H*. Thus, having $\hat{n},P$ parallel to the plates is the "hard flow" geometry, in fact the limit of hardest flow for uniform $\varphi(r)$ (HF limit), the only velocity response in the low frequency regime being the weak shear field written above $(G_{fl}^{planar})_\omega \approx i\omega[\alpha_3/\eta_2][(V_p)_\omega/V_{sat}] \equiv (G_{fl}^{HF})_\omega$, which we now refer to as $G_{fl}^{HF}$. With this weak velocity response the effective nematic reorientation viscosity coefficient seen by the electric field driving is $\approx \gamma_1$, [19,20,51] that for reorienting the director in a nematic fluid at rest, and therefore nearly the maximum $\gamma$ possible, the signature of a the HF limit state. The HF bulk $N_F$ viscosity $\gamma_{HF} = \gamma_1[1 - (\alpha_3/\eta_2)^2] \sim 0.92\gamma_1$, and resistance $R_{HF} = [\gamma_{HF}/P^2]d/A$ are large, as is the flutter dissipation, $D_{fl}^{HF}$.

If, on the other hand, the *homeotropic* geometry of **Fig. 6C** could be employed with a fluttering $N_F$, for example with an oscillating in-plane field, the induced flow gradient in the low frequency regime would be:

$$(G_{fl}^{EF})_\omega = i\omega[\alpha_2/\eta_1][(V_p)_\omega/V_{sat}] = [\alpha_2/\eta_1][\dot{\psi}_\omega], \quad (5)$$

~3x times larger than $(G_{fl}^{planar})_\omega$, and the effective viscosity being driven would be the bend viscosity $\gamma_{EF} = \gamma_1[1 - (\alpha_2/\eta_1)^2] \sim 0.36\gamma_1$, which is much smaller than $\gamma_1$, so giving a much smaller $R_{EF} = [\gamma_{EF}/P^2]d/A$, in fact the smallest $\gamma$ and $R$ obtainable by exploiting the reduction of stress by flow, and the smallest dissipation, $D_{fl}^{homeo}$. This would be in the EF limit, so we take $G_{fl}^{homeo}$ to be the EF flow gradient, $G_{fl}^{homeo} \equiv G_{fl}^{EF}$, giving a much larger induced velocity and much smaller dissipation, $D_{fl}^{EF}$ than the $N_F$ counterparts. The expressions in blue in **Eqs. 4b, 5** represent the limits of the effectiveness of ferroelectric nematic fluttering: *e.g.*, in the EF case, angular velocity deposits torque with inverse viscosity $1/\eta_1$, which is converted into linear velocity gradient with viscosity $\alpha_2$.

Now, let us consider the thought experiment in which we make it possible for the planar $\hat{n}, P$ in **Fig. 6B,C** to be free to rotate in the (u,z) plane 90° to the homeotropic orientation, while continuing to flutter. Returning to **Eqs. 3,** If we set ω to be in the low-frequency regime, then, since $D_{fl} \propto R \propto \gamma$, and since such a rotation will decrease the effective $\gamma$ by ~3x (estimate below) from $\gamma_{HF}$ to $\gamma_{EF}$, significantly decreasing dissipation, the rotation degree of freedom would experience a net average DC force toward the homeotropic: the system seeks EF for $\omega\tau_o < 1$. On the other hand, if $\omega\tau_o > 1$ we have $D_{fl} \propto 1/R \propto 1/\gamma$, and dissipation is reduced at high, rather than low, viscosity. Now the system seeks HF and the planar alignment would be the most stable. If $C$ is reduced the low frequency regime expands, where $D_{fl} \propto |V_p|^2 C^2 \gamma$, so that decreasing $C$ will require increasing $|V_p|$ to reach the threshold for getting the FFS.

This example points the way toward understanding the FFS textures as follows. For $\varphi(r)$ uniform the $v(z)$ velocity field, shown in red in **Fig. 1E**, creates a system of charge-/flow-stabilized coupled vector fields, with $\Gamma_{fl}$ normal to the (u,z) plane, and the $\hat{n}, P, v_u$ parallel to the (u,z) plane, with $v(z)$ parallel to $\hat{u}$ and the plates, and velocity gradient $G_{fl}^{HF}$. Experimentally, there is stable flow alignment in RM734, so $\hat{n}, P$ will be oriented at the $N_F$ flow-alignment (Leslie) angle $\psi_L = \tan^{-1}(\alpha_3/\alpha_2)^{1/2} \approx (\alpha_3/\alpha_2)^{1/2}$ [51,19,20], found to be in the range (1°< $\psi_L$ < 20°). Considering this as a way of estimating $\alpha_3$, in Helfrich's molecular model, $(\alpha_3/\alpha_2)^{1/2} \approx W/L$, $W$ and $L$ are respectively the molecular width and length [52], giving $\psi_L \sim 16°$ for RM734 ($W/L \approx 0.25$). But literature data shows that $\psi_L$ increases with decreasing alkyl tail length, reaching $\psi_L \sim 10°$ for the shortest tails, which is the case for the RM734 structure [53,54]. RM734 also exhibits a strong tendency to associate end-to-end [13], which, from experiments on main chain oligomers [55] also tends to increase $\psi_L$, which can reach $\psi_L \sim 20°$. On this basis we estimate $\psi_L$ to be relatively large ($\psi_L > 15°$) in RM734, making $\alpha_3/\alpha_2 > 0.07$.

We can now estimate the viscosity ratios $\alpha_2/\eta_1$ and $\alpha_3/\eta_2$ relevant to $G_{fl}^{HF}$, $G_{fl}^{EF}$, $\gamma_{HF}$, and $\gamma_{EF}$. For compact rod-like nematics the Miesowicz viscosities $\eta_1, \eta_2, \eta_3 > 0$ are in the order $\eta_1 > \eta_3 > \eta_2$, roughly with $\eta_1 \sim 5\eta_2$ and $\eta_3 \sim 2\eta_2$. The relations between viscosities and Leslie coefficients are shown in the Supplementary Information. We have $\alpha_3/\alpha_2 << 1$, so $\gamma_2 \approx \alpha_2$ and $\gamma_1 \approx -\alpha_2$, with $\alpha_2, \alpha_3 < 0$. The rotational viscosity of RM734 $\gamma_1 > 0$ is $\gamma_1 = -\alpha_2 = +0.5$ Pa-sec at $T = 100°C$ [14]. Also $\eta_1 - \eta_2 = -\gamma_2 \approx -\alpha_2$, so $\alpha_2/\eta_1 \approx -[\eta_1 - \eta_2]/\eta_1 \sim -0.8$, making $\eta_1 \approx 0.62$ Pa-sec. Taking $\alpha_3/\alpha_2 \sim 0.07$ then $\alpha_3/\eta_2 = (\alpha_2/\eta_1)(\alpha_3/\alpha_2)(\eta_1/\eta_2) \sim (-0.8)(0.07)(5) = -0.28$. The ratio $G_{fl}^{HF}/G_{fl}^{EF} = (\alpha_3/\eta_2)/(\alpha_2/\eta_1) = (\alpha_3/\alpha_2)(\eta_1/\eta_2) \sim 0.35$. $\gamma_{HF} = \gamma_1[1 - (\alpha_3/\eta_2)^2] \sim 0.92\gamma_1$, and $\gamma_{EF} = \gamma_1[1 - (\alpha_2/\eta_1)^2] \sim 0.36\gamma_1$. $\alpha_3 = 0.035$ Pa-sec.

*Stability of $\hat{n}, P, v_u$* – We now consider the stability of the uniform $\varphi$ arrangement with $v_u$ parallel to $\hat{n}, P$, by exploring various deformations by external forces, first in which $v_u$ is rotated through an angle $\delta\varphi_{vn}$ about $\hat{z}$ out of the plane containing $\hat{n}, P$. In this reorientated state a viscous pressure gradient normal to the $(u,z)$ plane appears ([20], Fig. B.III.16]), which applies a torque density to $\hat{n}, P$ about $\hat{z}$, given by

$(\mathcal{T}_{fl})_\omega = \hat{z}[(\eta_3-\eta_2)(G_{fl}^{HF})_\omega]\sin\varphi_{vn}\cos\varphi_{vn} \approx \hat{z}[(\eta_3-\eta_2)(G_{fl}^{HF})_\omega]\delta\varphi_{vn} = \hat{z}[(\eta_3-\eta_2)][\alpha_3/\eta_2][\dot{\psi}_\omega]\delta\varphi_{vn} \sim [\alpha_3\dot{\psi}_\omega]\delta\varphi_{vn}$

making $c\mathcal{T}_{fl} = [\alpha_3\dot{\psi}_\omega]$ a flow-driven effective bulk orientational anchoring coefficient that strongly couples $v_u$ to $\hat{n}, P$. This torque acts to restore the parallel state of $\hat{n}, P, v_u$, with an extensive component of the nematic flow gradient $G_{fl}^{NF}$ in the $(t,u)$ plane tending to stretch the molecules lengthwise and rotate them toward being parallel to the $(u,z)$ shear plane. Because the $\hat{n}, P$ flow-established alignment in the $(t,u)$ plane will change sign when the oscillating $G_{fl}$ changes sign, this stretching by $\Gamma_z$ takes place for both signs of velocity in the drive cycle, effectively rectifying the alternating AC velocity to give a spring-like average restoring torque density keeping $\hat{n}, P$ along $v_u$. If $v_u$ were channeled by the cell then such a restoring torque would act like an anisotropic boundary condition providing uniform azimuthal director orientation. The effect of such a torque can be assessed by comparing typical LC Rapini-Papoular surface anchoring energy coefficients ($10^{-7} < \mathcal{T}_{RP} < 10^{-4}$ J/m$^2$) with the effective surface anchoring torque/area due to fluttering, $|\mathcal{T}_{fl}|d \sim PV_{sat} \sim 0.3$ J/m$^2$ obtained from $\mathcal{T}_{fl}$. This large ratio, $\mathcal{T}_{fl}/\mathcal{T}_{RP} \gtrsim 10^4$, accounts for the suppression of the random planar textures in *Figs. 1,5* by fluttering.

*Dissipation of planar fluttering states in the low frequency regime* – In our cells the applied AC voltage oscillates about $V(t) = 0$, making the oscillating applied field along $z$, the average director orientation planar, with average block polarization orientation $\langle\psi(t)\rangle = 0$. Thus, in the discussion that follows, we consider only *the planar* homogeneous fluttering $N_F$ shown in *Figs. 1G,1H,6B,7C*. We analyze our data from the point of view of using ferro-electro-nemato-hydrodynamics to directly probe how manipulation of dissipation in the fluttering ferroelectric nematic leads to the effective elastic behavior and interfacial structures evident in the FFS textures.

In the uniform $\varphi(r)$ condition the HF FFS *planar* state is locked-in by the plates and the translational symmetry along $\hat{u}$. We now address textural modifications and features whereby the dissipation of this uniform HF state can be altered. Such modifications can be divided into classes of continuously-broken and discretely-broken translational symmetry, the former comprising the director deformations (splay, bend, and twist) of uniaxial nematics, and the latter comprising interfaces, defect cores and domain boundaries.

Another binary classification of HF FFS *planar* phenomena are those that show up in the FFS textures; and those that do not, that is to say those that may lower dissipation and those that may raise it. For example, point defects with large $\nabla \cdot P$ and interfaces where $P$ has a component normal to domain boundaries can be found, splay can fill entire FFS areas, but bend and twist are expelled.

We consider the effect of continuous splay, bend, or twist deformation on dissipation by the uniform state. The bend and splay deformations are two dimensional (2D) $\varphi(x,y)$ fields with $+2\pi$ topological defects where $\hat{n},P$ is respectively either normal (splay) or tangential (bend) to circles centered on the defect core. The twist deformation is $(x,y)$-independent, where $\varphi$ varies linearly with $z$. These considerations are all in the low frequency regime.

### *Continuous translational symmetry breaking in FFS textures*

(*i*) <u>bend deformation of $\varphi(x,y)$</u> (*not observed in FFS textures reported here*) – Bend is the favored textural feature in the equilibrium $N_F$ phase, in particular in thin films which are boundary torque-free, such as in planar $N_F$ films between slippery solid or isotropic liquid and/or air surfaces [32,33]. Remarkably, such structures and other bend geometries are completely absent in the FFS textures, another "conjugate" equilibrium/FFS texture feature which we propose to explain as follows. Imagining a $2\pi$ bend defect in the $\hat{n},P,v_u$ field, having a local $\hat{u}$ coordinate tangent to the concentric circles centered on its core, the $\hat{n},P,v_u$ field variables satisfy periodic boundary conditions on these circles, with an orientational period of $2\pi$, and are thus effectively translationally symmetric along $\hat{u}$. Being so, in the defect structure $\hat{n},P,v_u$ would be locally in a uniform $\varphi$ mode, with the local flutter generated velocity gradient $G_{fl}^{HF}$ everywhere, and $v_u = G_{fl}^{HF} d/2$. In the presence of bend this flow generates in-plane vorticity $\omega_z = \nabla \times v = (1/\rho) \partial(\rho v_u)/\partial \rho = v_u/\rho$, where the defect center is $\rho = 0$ and the local bend magnitude $B \equiv \partial \varphi/\partial u = 1/\rho$. However, more generally, in a complex texture where there is bend deformation, then locally $\omega_z = v_u B$, with $B^{-1}$ the local bend radius of curvature of $\hat{n}(r)$ field. The vorticity appears as in-plane shear flow with gradient $\omega_z = v_u B$, giving bend a dissipation per unit volume from *Eq. 4B*:

$$D_{\text{bend}} = \tfrac{1}{2}\eta_2 \{\tfrac{1}{2}d(\alpha_3/\eta_2)(\omega|V_p|_\omega/V_{\text{sat}})\}^2 \, B^2 \qquad (6a)$$

(*ii*) <u>continuous twist deformation</u> (*expelled at the transition to the FFS textures*) – In analogy to bend, starting from the uniform $\hat{n},P,v_u$ state, twist deformation along the a z-axis generates vorticity $\omega_y = \partial(v_u)_y/\partial z = [G_{fl}^{HF} d/2]T$, where $T^{-1} = \partial\varphi/\partial z$ is the spatial rotation rate of $\hat{n},P,v_u$ along z, giving *Eq. 6b*:

$$D_{\text{twist}} = \tfrac{1}{2}\eta_3 \{\tfrac{1}{2}d(\alpha_3/\eta_2)(\omega|V_p|_\omega/V_{\text{sat}})\}^2 \, T^2 \qquad (6b)$$

(*iv*) <u>continuous splay deformation of $\varphi(x,y)$</u> (*orange to yellow shading in Figs. 2,5,7*) – Weakly-broken translational symmetry along $\hat{n}$ generates easy flow in FFS textures. DTOM shows that the $\hat{n}(\rho)$ in the +2π topological defect is radial and $P(\rho)$ is directed either toward or away from the core, as shown in *Figs. 2,4,5*. Apart from the core, where translational symmetry along $\hat{n}$ is strongly broken, to be discussed in the next section, this ubiquitous FFS texture textural feature has the director uniform in z and splayed in (x,y), with each place locally obtainable from the HF state by a continuous deformation, suggesting that splay may reduce dissipation and therefore require EF. The radial cross-section of a 2π defect along a line cutting it in half is sketched in *Figs. 7B-E*. The in-plane component of velocity, $v_\rho$, generated by HF flutter is radial, and of uniform average magnitude, $v_\rho = v_{\text{HF}} = G_{fl}^{HF} d/2$. However, such a velocity field is not divergence-free, as then $\nabla \cdot v_\rho(\rho,z) = v_{\text{HF}}(\rho,z)/\rho$, violating the fluid incompressiblity condition. This can be avoided in the overall radial flow by assuming the upper-in and lower-out flow structure in the three-block pattern of *Fig. 1H*, However, just considering the flow in the central block, there is still nonzero $\nabla \cdot v_\rho(\rho,z)$ within the upper and lower halves. This requires us to add a velocity component $v_z(\rho,z)$ such that $\partial v_z(\rho,z)/\partial z = -v_{\text{HF}}(\rho,z)/\rho$, in which case flow in the *z* direction will supply the outflow (inflow) of upper-half fluid, and to carry off the excess inflow (outflow) of the lower half on alternating cycles.

This scenario is in fact promoted by EF pumping because, as the director field is deformed from uniform to fan-shape, translational symmetry along $\hat{n}$ is lost in a continuous fashion, as can be seen if we consider a typical (d$\rho$ x $\rho$d$\varphi$ x d) volume element of fluid in the radial field. Each such element is subject to an unbalance of forces along z , as sketched in *Fig. 7D*, from the mutually out-of-phase force applied by adjacent elements neighboring on its inner and outer surfaces. As the $\hat{n},P,v_u$ field is splayed these surfaces develop different area, an imbalance which applies a net fluttering stress $\sigma_z(\rho,z)$ along z to the element. This stress drives the z component of flow *toward* maintaining the $\nabla \cdot v = 0$ condition. Importantly this driving force is in an EF geometry, with the result that, starting from the uniform $\varphi(r)$ state, dissipation will be reduced by introducing splay, meaning that the FFS texture ground state is splayed. This accounts for the observed preponderance of splay in the FFS textures at the expense of bend and twist, and introduces the question of what the preferred magnitude of splay deformation is. The $+2\pi$ defects may serve as laboratories for addressing this question, since they present a range of splay magnitudes in their radial structure, down to a $\rho$ value where $v_z(\rho)$ couples to a toroidal roll in the core, discussed below, a cutoff radius perhaps at the scale of preferred splay.

*__Discrete translational symmetry breaking generates easy flow in FFS textures__* – Here we analyze features of the textures where translational symmetry along $\hat{u}$ is discretely broken, enabling the system to locally reduce dissipation by introducing planar EF regions. This is achievable by forcing $v_u$ to zero at $N_F$ boundaries, and in defect cores, and by forcing $v_u$ to change magnitude as in dissipation stabilized kink (DSK) domain boundaries, as follows:

(*v*) <u>bubble boundaries </u>(*magenta lines in Figs. 2,5,7*) – The clearest breaks of the translational symmetry along $\hat{u}$ are termination of $\hat{n}$,*P* at the LC/bubble boundary, where ***Fig. 2C*** shows $\hat{n}$,*P* approaching the LC/bubble interface along a line (taken to be along *x*) at normal incidence to the interface line, along *y*, a structure shown in section in ***Fig. 6D***. At the boundary ($x = 0$, with $\hat{u} = \hat{x}$), because of incompressibility, *v* must follow the fluid surface, turning to the *z* direction to pass fluid from the upper to the lower half-sheet (***Fig. 6D***). On the air side of the $x = 0$ surface there is no fluid to counterflow and impede vertical flow for $x < 0$ as in the 2D bulk. Additionally, this vertical flow $v_z(x < 0)$ at the end is now in the EF orientation for efficient pumping by flutter, generating a pressure gradient $\partial p/\partial z$ set in magnitude by $G_{fl}^{EF}$ that transports more fluid up or down than can be supplied or removed away from the boundary, set in magnitude by $G_{fl}^{HF}$. This excess flow leads to the formation of a boundary layer of width $\lambda$ along *x* in the form of a velocity roll at the interface, with roll axis parallel to the boundary. In this layer, if $v_z(x = 0)$ is instantaneously positive, then, for increasing distance from the interface, $v_z(x)$ will decrease, change sign at $x \sim -d$, and then decrease exponentially to zero with a decay length $\lambda$, over which $v_x(x)$ will relax to its (small) bulk HF value. Within the boundary layer we have $(\partial v_x(x,z)/\partial x) \propto p(x,z)$ due to the pressure driven leakage through the cell midplane, and $(\partial p(x,z)/\partial x) \propto v_x(x)$ from the laminar flow along *x* in opposite directions across the cell midplane, giving exponential decay. In the roll both the up and down flows are in the EF geometry, also pumping a $v_x$ of magnitude $G_{fl}^{EF}$ parallel to the plates.

If we now consider a situation where $\hat{n}$,*P* approach the boundary at an angle to the interface, say 45° from the interface line, we can express the driving torque $\Gamma_{fl}$ as a sum of components normal and parallel to this line ($\Gamma_{fl} = \Gamma_x + \Gamma_y$). The normal component $\Gamma_x$ drives flow parallel to the interface, a direction which maintains translational symmetry, generating $G_{fl}^{HF}$ shear, $\partial v_y(z)/\partial z$, parallel to the interface line. However, the parallel component of this torque, $\Gamma_y$ drives $G_{fl}^{EF}$ shear as in the normal case and, thereby, the roll in the (x,z) plane, having the large EF velocity at the interface line, the EF condition giving it a large $v_x$ normal to the interface, thereby turning $v$ toward $\hat{x}$. As discussed in the estimate of $\mathcal{T}_{fl}$ above, this velocity reorientation will apply torque $\Gamma_z$ to $\hat{n}, P$, reorienting $\hat{n}, P$ toward also being normal to the interface line. This establishes an interface geometry that has $\hat{n}, P$ preferentially aligned normal to the interface to give minimum dissipation, with an anchoring strength that is enhanced by the EF contribution to $v_u$ in the roll, and is apparently large enough to overcome the electrostatic preference that $\hat{n}, P$ be parallel to the interface line. Thus, the flutter mechanism can account for this unusual boundary orientation of $\hat{n}, P$.

(*vi*) <u>*defect cores*</u> (*red shading in Figs. 2,5,7*) – Breaks of the translational symmetry along $\hat{u}$ also occur at the +2π defect cores, where *P* changes sign upon crossing the defect axis, as sketched in **Fig. 6B**, which is a section on a plane normal to the plates and cutting through the core. The forced sign reversal of *P* at the axis generates nonzero $\partial v_x/\partial x$ and therefore $\partial \sigma_{zx}/\partial x$, which drives easy flow $v_z(x)$ in the core a net flow creates counter flow of $v_u$ in the radial direction at the core, which, according to discussion (*i*), will lead to $v_z$ flow along the defect axis. The cylindrically symmetric core volume will transform from HF to EF, producing a toroidal roll in the core region, centered on the defect axis and having a large radial $v$, which, as at the bubble boundary, establish a radial boundary condition on $\hat{n}, P$ around the core.

(*vi*) <u>*dissipation-stabilized splay-bend kink (DSK) domain boundaries*</u> (*cyan lines in Figs. 2,5,7*) – In order to analyze the 1D splay-bend wall, indicated by cyan lines in **Figs.2,4,5,7A** and sketched in cross-section in **Fig. 7F**, it is desirable to recall its equilibrium cousin, the electrostatically stabilized *polarization stabilized kink* (PSK) splay-bend lines found in equilibrium fluid ferroelectric liquid crystals [26,56,57,58,59,60]. In these equilibrium systems splay of the *P*(*r*) field produces volume space charge, $\rho_P(x) = -\nabla \cdot P$. As a result, in order to avoid the resulting electrostatic energy, 2D and 3D textures of high *P* LC materials tend to have $\nabla \cdot \hat{n}$ and $\nabla \cdot P \approx 0$ everywhere, exhibiting only bend deformation in the bulk. A typical bend-textural motif of *P*(*r*) in 2D is to anneal to a finite length scale that leaves a population of +2π topological defects, in which *P* is everywhere tangent to circles concentric with the core (pure bend) [59,60]. Boundaries between neighboring defects, and the compensating population of -2π topological defects require PSKs, linear splay-bend defect lines stabilized by the internal splay of *P*(*r*) [26,57,58,59]. This can be understood by considering the geometry of *P*(*r*) in **Fig. 7F**, to be that of an equilibrium splay-bend defect line along $\hat{y}$, in which case deposited space charge $\rho_P(x) = -\partial P_x/\partial x$ is of opposite sign on opposite sides of the wall, generating an attractive force across the line that confines it [57,58]. Additionally, it is impossible to make a -2π defect in a 2D vector field of constant magnitude without splay, so if *P* is large this splay is confined to four mutually perpendicular PSK defect lines meeting at the -2π defect core, with domains of uniform **P** in between, $P_y$ switching sign as a PSK line is crossed, but with $P_x$ the same on both sides [57,58]. These PSK lines are features of excess energy and so would disappear if they could. They cannot, but they are the minimum-energy structural features that enable the formation of large low-energy areas free of splay, and arrays of +2π bend defects on larger scale.

Thus the splay-bend lines of *Fig. 7F* play a similar role in the realm of the dissipative FFN, and will be termed dissipation-stabilized kinks (DSKs). In the dissipative FFS texture the dominant EF nature of the core of the +2π defects and their dissipation-favored splayed $\hat{n}$,**P** field, lead to +2π defects covering most of the area of the FFS texture, but between neighboring +2π defects are distinguishable line defects, shown emerging in *Fig. 1E*, and in full flower in the 2D defect lattice in *Fig. 5*. Of note in these images is that the splayed $\hat{n}$,**P** field around each core persists in its nearly undeformed radial geometry out to close to its bounding square.

*Fig. 7F* is the projection of a splay-bend onto a plane normal to the wall, showing $\hat{n}$,**P** in the FFS texture areas approaching the wall, on opposite sides making angles ±β with respect to the *x* axis, the normal to the wall, projecting, for example, $P\cos\beta$ onto $\hat{x}$. Within the wall, near its center, at *x* = 0, we have β = 0, so that this projection along $\hat{x}$ is $P_x$ =*P*. The resulting nonzero ∂$P_x$/∂x breaks translational symmetry along $\hat{x}$, generating nonzero ∂$v_x$/∂x and therefore ∂$σ_{zx}$/∂x, which drives easy flow $v_z(x)$, where $v_z(x) \propto (1 - \cos\beta)G_{fl}$ With an opposite gradient on the other side of the wall, $v_z(x)$ of opposite sign forms an EF roll, shaded cyan in *Fig. 7E*.

*Effective mean elastic and interface interactions due to dissipative forces* – In AC dissipative systems such as bird flight, the rowboat of *Fig. 9*, or the fluttering $N_F$, forces generating net motion arise from the differences between forward and reverse strokes. For example, a bird wing is articulated such that it takes on a bent shape on the upstroke to reduce $P_u$, the downward momentum transferred from the air, and a flatter shape on the downstroke to maximize $P_d$, the upward momentum transferred from the air, producing net lift. A dimensionless factor, $0 < f = |(P_d - P_u)/P_d| < 1$, characterizes the range of possibilities, where the lower limit $f = 0$ corresponds to identical forward and reverse strokes, e.g. a rowboat with stem-to stern mirror symmetry, and symmetric paddling. In the $N_F$ case, the small-ψ linear AC dynamic model of *Eqs. 2-6*, fluttering does not produce net time averaged DC interactions because in the liner regime the forces generated on the phase space trajectory, $\Gamma_{fl}(\psi(t))$ for ψ(t) going from 0° to 90°cancel those going from 90° to 0°. Consistent with this condition is the observation that the threshold for generating the FFS textures is $V_p$ being comparable to $V_{sat}$, where the $\psi(t)$ response is distinctly nonlinear, saturating at $|\psi(t)| \sim 90°$. In the $N_F$ case, pursuing the bird wing analogy, we can use the linear analysis in the sections above to estimate flow generated contributions to elasticity and interface interactions, the flow equivalent of $P_d \approx P_u$ ($f = 0$) in the linear regime of the bird problem, but the equivalent of the calculation of $f$ will require solution of the nonlinear nematoelectrohydrodynamic equations for the various interactions and geometries considered, which will not be pursued here. Rather we summarize below the various half-cycle dissipative forces in the linear regime, recognizing that for each example the time-averaged contribution will be only a fraction of this value.

But first we consider are the expected values of $f$. We provide examples here of the entire range of $f$ from 0 to 1. Thus **Fig. 8** shows that $f \approx 1$ is possible, the effects of consecutive forward and reverse strokes being completely different: the first stroke generates a dramatic change from a defected to an ordered pattern, while the reverse stroke changes the result very little beyond that. This happens because this particular written pattern can be stabilized by a series of the same alternating strokes. Once so written, identical subsequent strokes of either sign have very little effect on the pattern: thus the same system achieves a particular $f = 0$ condition for a particular choice of backward / forward stroke sequences. The evolution of a pattern upon changing from stroke duration of 0.5 sec to stroke duration of 2.5 sec is shown in **Fig. 6**. Changes in the pattern from stroke to stroke are visualized and occur entirely when $|V(t)| \sim V_{sat}$: the nonlinear regime is required. The beginning and ending lattices are both stable, but with very different, self-selected lattice cell dimensions.

The fluttering, of uniform magnitude $\Gamma_{fl}$, produces a bulk dissipation $D = \frac{1}{2}\Gamma_{fl}\dot{\psi}^* = \frac{1}{2}\gamma_{HF}\dot{\psi}\dot{\psi}^*$. This, in turn generates a bulk torque/volume about $z$, of magnitude $\mathcal{T}_{fl}$, acting to keep $v_u$ parallel to $\hat{n},P$ fills the $N_F$, where:

$$\mathcal{T}_{fl} = \hat{z}\alpha_3\dot{\psi}, \qquad (\mathcal{T}_{fl})_z \sim (\alpha_3/\gamma_1)\gamma_1\dot{\psi} \sim (\alpha_3/\gamma_1)PV_{sat}/d, \qquad (7)$$

and $\gamma_{HF} \approx \gamma_1$ is the hard flow viscosity obtained above. $(\mathcal{T}_{fl})_z$ couples $v_u$ to $\hat{n},P$ enabling flow to affect texture.

Typical Rapini-Papoular coefficients of dimension (energy/area) for LC surface anchoring are in the range ($10^{-7} < A_{RP} < 10^{-4}$ J/m$^2$), so an air bubble/$N_F$ boundary in a cell of thickness $d$ will have anchoring energy/length ($10^{-13} < A_{RP}d < 10^{-11}$ J/m). According to **Fig. 10** and its text, near an air-$N_F$ boundary of length $L$, $\gamma$ is reduced by easy flow in a volume $Ld^2$ along the interface, producing an effective Rapini-Papoular interface anchoring torque per unit length

$$A_{RP}d = \boldsymbol{\mathcal{T}}_{fl}d^2 = \hat{z}(\gamma_{HF} - \gamma_{EF})\dot{\psi}\,d^2 \sim PV_{sat}d,$$

with $A_{RP} \sim PV_{sat} \sim 0.3$ J/m² dominating typical equilibrium values. It was pointed out that the fluttering deposits energy at a high rate relative to that in a cell's elastic deformation, which can act as effective elasticity of the $\hat{n}, \boldsymbol{P}$ field. The comparison of torque/area, $K(\pi/d) \sim 3\times10^{-5}$ J/m², typically transmitted by the Frank elastic molecular field with the field-applied ferroelectric value, $PV_{sat} \sim 0.3$ J/m², enables estimation of the energy scale of an effective flow-based elasticity through elastic constant $K_{eff}$:

$$K_{eff} \sim PV_{sat}\,d \sim \gamma_{eff}\dot{\psi}\,d^2,$$

Where $K_{eff}$ can be ~$10^4$ times Frank values, which accounts for the effectiveness of flow at definitively eliminating equilibrium textures and establishing FFS textures. Correspondingly the bulk dissipation $D = \tfrac{1}{2}\Gamma_{fl}\dot{\psi} = \tfrac{1}{2}\gamma_{eff}\dot{\psi}^2$ can be related to the effective elasticity.

$$D \sim \dot{\psi}(K_{eff}/d^2).$$

with the effective surface anchoring torque/area due to fluttering, $|\mathcal{T}_{fl}|d \sim PV_{sat} \sim 0.3$ J/m² obtained from $\mathcal{T}_{fl}$. This large ratio, $\mathcal{T}_{fl}/\mathcal{T}_{RP} \gtrsim 10^4$, accounts for the suppression of equilibrium planar textures in *Figs. 1,5* by fluttering.

*The equilibrium/active nematic FFS transition* – This FFS texture state was reached by a discrete transition from the equilibrium condition once the drive is initially applied. The simplest picture would have the equilibrium as a state of even higher dissipation that the HF FFS texture. One possibility for this is that for weak drive the sticky boundary conditions at the electrodes suppress flow altogether, and that the initial transition is to the multiblock hard flow states of *Figs. 1G,H*.

*MATERIALS AND METHODS*

*Electro-optics* – For making electro-optical measurements, The mixtures were filled into planar-aligned, in-plane switching test cells with either unbuffed uncoated electrodes or ones coated with alignment layers and unidirectionally buffed parallel on the two plates, which were uniformly separated by $d$ either $d = 0.8$ μm or $d = 8$ μm.. Such surfaces give a quadrupolar alignment of the N director along the buffing axis, and polar alignment of the $N_F$ on each plate, the latter making cells having uniform director/polarization field parallel to the plates and buffing direction [34].

*ACKNOWLEDGEMENTS*

This work was supported by NSF Condensed Matter Physics Grants DMR 1710711 and DMR 2005170, and by Materials Research Science and Engineering Center (MRSEC) Grant DMR 1420736.

# FIGURES

*Figure 1*: Electrostatics of producing polarization/director flutter in a ferroelectric nematic. (*A-C*) The ferroelectric nematic is uniaxial with director $\hat{n}$ and polarization *P* mutually parallel and along the uniaxis. The basic geometry of "block polarization switching" [21,24] is shown, in which polarization charge self-energy maintains a uniform *P* field in the $N_F$ fluid. Ferroelectric LCs generally have a dielectric layer at the electrode surface, here of capacitance/area, *c*. For slowly varying voltage, *V(t)*, applied to the electrodes, with *V(t)* smaller than the saturation value $V_{sat} = P/c$, the electric field in the LC can be cancelled by the orientation of *P*, given by $\psi$, in *Eq. 1*, establishing the electrostatically stabilized relationship $\sin\psi(t) = V(t)/V_{sat}$. (*D,F,G*) An AC voltage $V(t) < V_{sat}$ drives flutter of $\hat{n}$,*P*, about $\psi = 0$, applying torque density, $\Gamma$, to the fluid, locally normal to $\hat{n}$,*P*, with the resulting stress generating flow of the LC, *v(r)*, in the plane

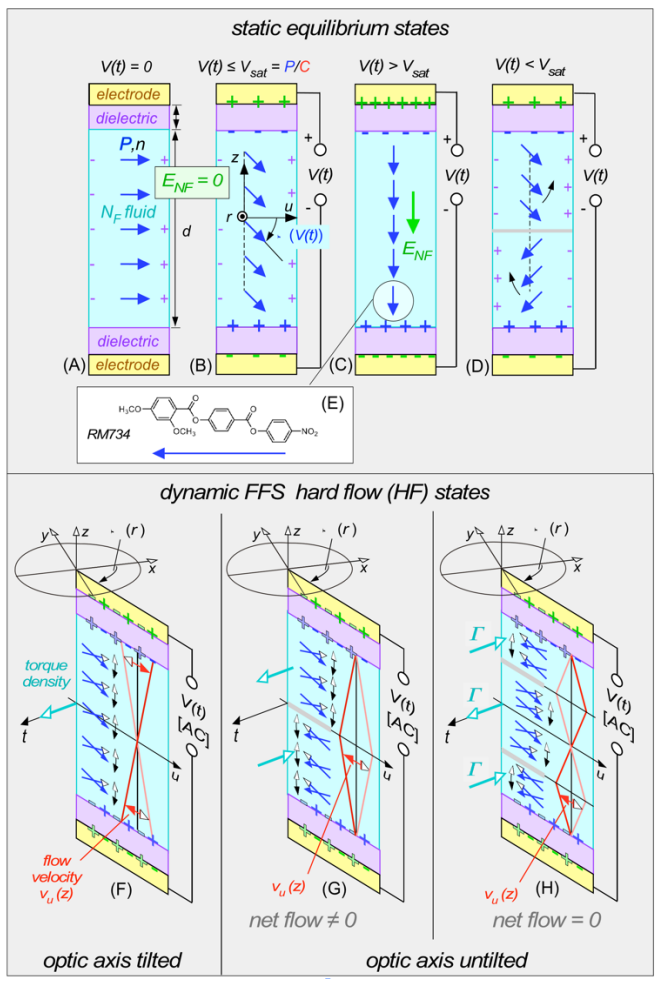

normal to $\Gamma$. The reorientation pattern and resulting flow will take the form of a single block as in (*B, F*) if the LC fluid is free to slip at the surfaces, or multiple blocks rotating in opposite directions, as in (*D,G,H*) if *v(r)* = 0 at the surfaces. (*G*) gives a net flow along *u*, while (*H*) is obtained if net flow along *u* is blocked. For small amplitude $\psi(t)$ this planar geometry forces *v(r)* parallel to $\hat{n}$, which is the is the highest-dissipation, lowest efficiency, lowest flow geometry of flutter-induced fluid motion [hard-flow (HF) limit]. (*E*) The molecule studied [10], which has a longitudinal electric dipole moment of magnitude ~ 11Debye.

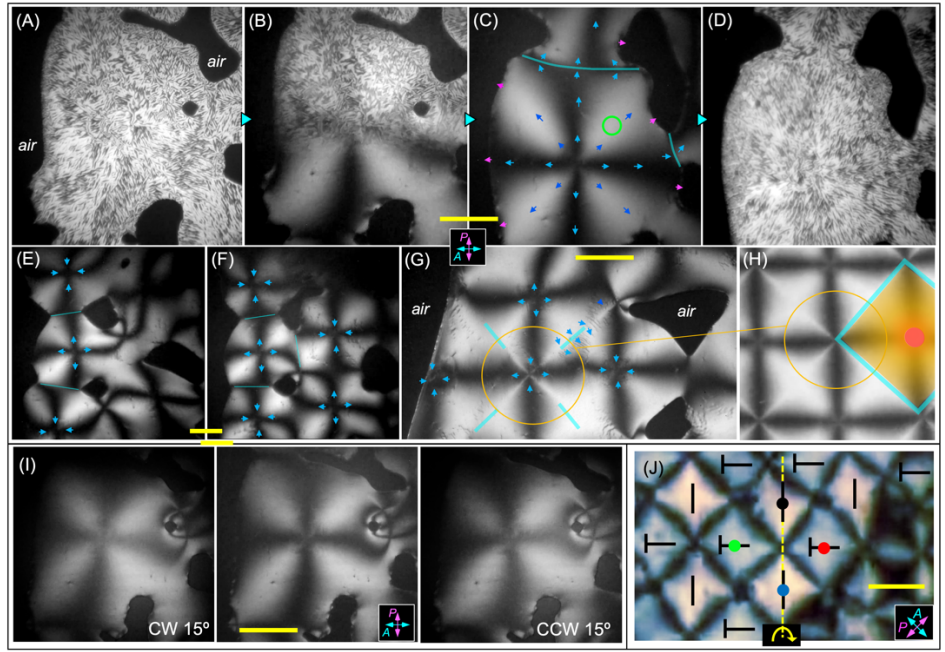

*Figure 2*: Random planar and Fluttering Ferroelectric nematic Smooth (FFS) textures in a bare ITO, $d = 0.8\mu m$ RM734 cell at $T = 100$ ºC. (*A*) Initial $V(t) = 0$ random planar texture. (*B*) Coexistence of random planar and FFS textures at $V_p = 10V$ and a drive frequency $f = 700$ Hz. The frame rate is 60 Hz so that each of the images is the time average of instantaneous images over ~8 cycles. (*C*) FFS texture as grown from the random planar with increasing peak voltage $V_p$. Light and dark blue arrows indicate the 2D bulk direction of *n̂,P*, while magenta arrows indicate *n̂,P*, direction at the NF/air interfaces. The defect is a +2π singularity with a radial *n̂,P* field. The cyan line is a Dissipation Stabilized Kink (DSK) domain boundary between the radial fields of adjacent +2π defects. (*D*) Reappearance of the random planar texture upon returning to $V(t) = 0$. (*E-G*) Larger area showing control of the organization of the 2π defects by their mutual interaction and by bubbles. (*C,E-G*) Any area in these FFS textures can be extinguished by appropriate sample rotation, indicating that they have little or no twist, but rather uniform alignment of *n* along *z*. Nascent ordered array showing DSK borders between +2π defects. DSK lines can from a cross to generate -2π defects complementary to the +2π defects (gold circle) (*H*) Area of the periodic defect lattice in a fresh cell with no bubbles observed at $f = 900$ Hz and $V_p=5V$, showing the similarity to arrangements of +2π defects when they are first appearing. The colored plaquette shows the elemental unit that tiles to form the lattice. (*I*) Uncrossing the analyzer by 15º in right (CW) and left (CCW) directions produces similar patterns, showing that there is negligible twist in these FFS textures. (*J*) Tilted cell experiment, showing where the four $I_\varphi(t)$ curves of *Fig. 3B* ( • • • • ) for the full range of $\psi(t)$ were obtained, at four places on a circle centered on a radial +1 defect, having different $\varphi(x,y)$ values relative to the polarizer [45º,135º, 225º, 315º]. The cell is tilted through ~25º about the yellow dashed line tilt rotation axis, with the image, showing two distinct pairs of quadrants with similar birefringence color, the combination which indicates that the local average *n̂,P* orientation is always parallel to the cell plane for the full range of $\psi(t)$, as in *Fig. 1F,G*, giving the director tilt with the usual "T" notation. Scale bars = 100μm.

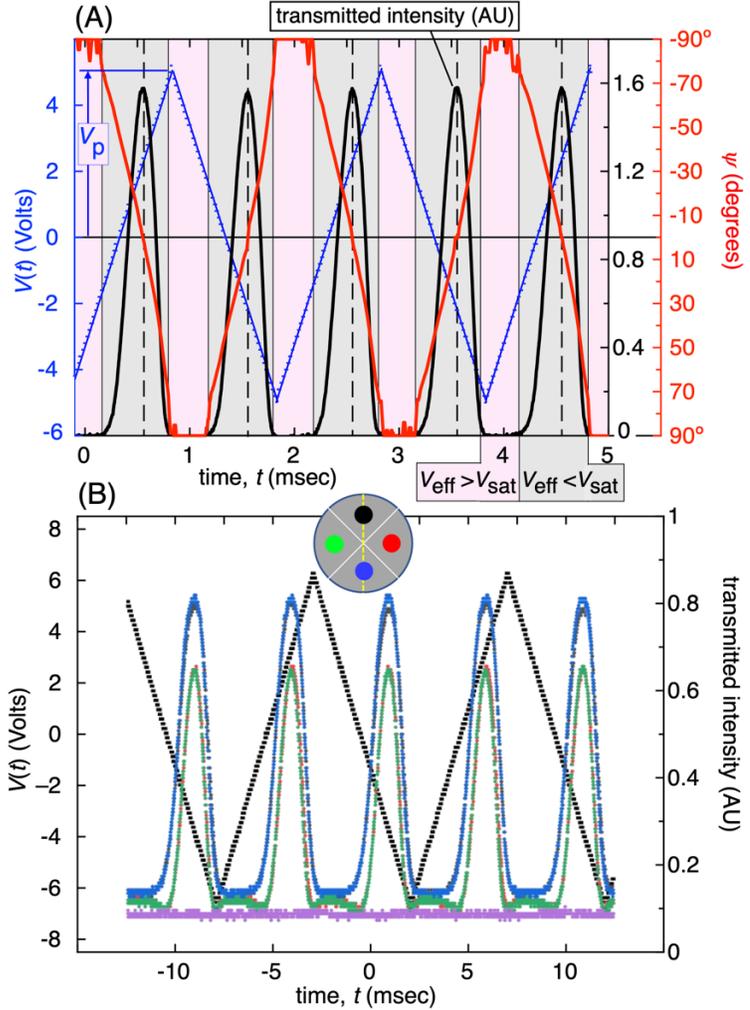

*Figure 3*: Electro-optic and electro-orientation response of a bare ITO - RM734 cells at $T = 100$ ºC, exhibiting the essential features and applicability of the of "block polarization switching" model for the field-induced fluttering drive of a planar $N_F$ cell, at $f = 500$ Hz.

(***A***) <u>Normal optical incidence</u> on a $d = 0.8 \mu m$, $V_{sat} = 3.5V$ cell. $V(t)$ (blue curve) is the voltage applied to the cell, in this case a triangle wave of pek voltage $V_p = 5V$, and $\psi(t)$ (red curve) is the rotation of $\hat{n},P$ from the planar orientation. Laser light is weakly focused to a ~70 μm diameter spot through crossed polarizer/analyzer with $\hat{n},P$ oriented at 45º to the optical polarization, giving maximum transmitted intensity, *I* (black curve), at $V(t) = 0$. Assuming the NF to make a uniform uniaxial birefringent slab of($\psi$). In the "block polarization" mode $\psi$ is uniform along z in the cell except for nanoscale-thickness layers at the surfaces, so , given birefringence $\Delta n = 0.19$, $\psi$ can be calculated from *I*, giving $\psi(t)$. (***B***) <u>Oblique optical incidence</u> on a $d = 1.0$ μm, $V_{sat} = 4.5V$ cell, showing where the four $I_\varphi(t)$ curves of ***Fig. 3B*** (● ● ● ●) for the full range of $\psi(t)$ were obtained, at four places on a circle centered on a radial +1 defect, having different $\varphi(x,y)$ values relative to the polarizer [45º,135º, 225º, 315º], as shown in ***Fig. 1J***. The cell is tilted through ~25º about the yellow dashed line tilt rotation axis, so that we should always have $I_{135}(t) = I_{315}(t)$, as is found in (B). If $\hat{n},P$ is not tilted from the cell plane then there will be two distinct, like pairs, $I_{135}(t) = I_{315}(t)$ and $I_{45}(t) = I_{225}(t)$, also as found in (B). If $\hat{n},P$ is tilted on average from the cell plane then we would have $I_{45}(t) \neq I_{225}(t)$, not observed in (B). We conclude that the local $\hat{n},P$ profile is on average untilted, as in ***Figs. 1F,G***. For $\psi(t)$ saturated at $|\psi| \approx 90º$ the nearly homeotropic remnant birefringence is nearly the same in all four quadrants. (***A,B***) These cells have some resistive leakage current through the interfacial capacitors, which puts some free charge on the LC/interfacial capacitor interfaces. This charging shifts the zero crossings of $\psi(t)$ to ~0.3 ms period behind those of $V(t) = 0$.

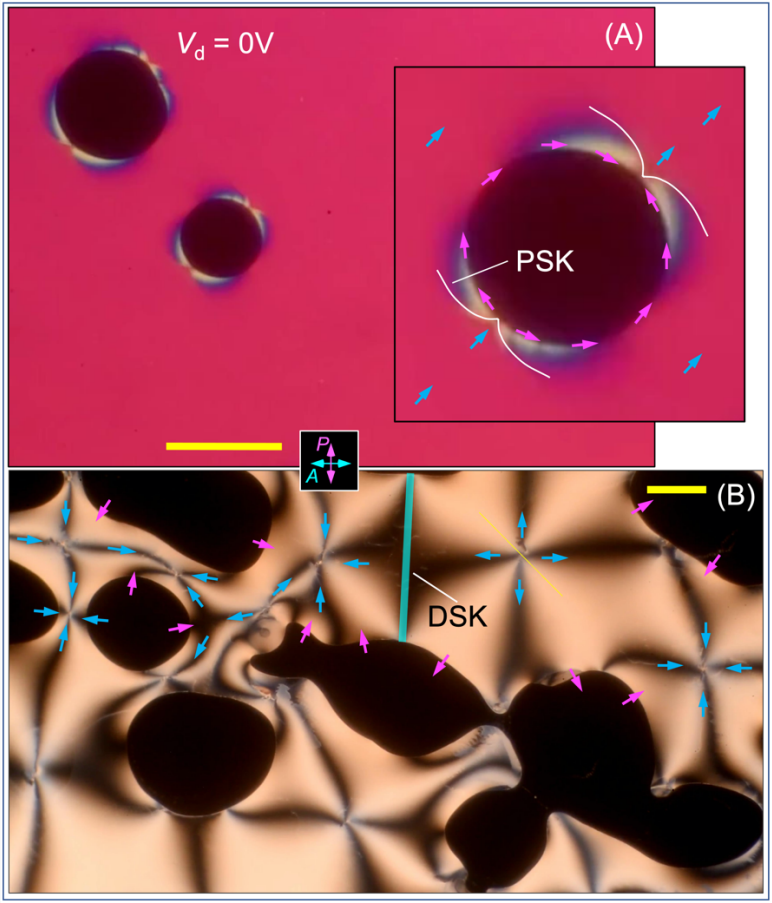

*Figure 4*: Textures obtained in a rubbed polymer ITO, $d = 2.0$ μm RM734 cell at $T = 100$ °C. (*A*) Initial planar-aligned $N_F$ monodomain with several bubbles. Bubbles in an aligned texture generate polarization stabilized kinks (PSKs), parabolic lines where the normal component of ***P*** is continuous and the parallel component changes sign, mediating the transition between the aligned (cyan) and tangential to the bubble (magenta) orientations [56]. (*B*) With $f = 1000$ Hz, $V_p = 10$ V, after the transition to the fluttering state the driven FFS texture exhibits no evidence for influence by the surface alignment. All vestiges of texture (*A*) are gone, including the tangential alignment at the bubble/LC interfaces, which is now normal in the FFS state (magenta arrows). The cyan line in (*B*), the boundary between $+2\pi$ defects of opposite sign of ***P*** is a Dissipation Stabilized Kink (DSK), the domain boundary structural equivalent to the equilibrium PSK in (*A*), enabling the formation of arrays of neighboring $+2\pi$ defects as in the arrays of **Figs. 2,7** and the periodic lattice of **Fig. 5**. Tilting of the sample plane about the yellow line in (*B*) shows that the local average optic axis in this texture is always parallel to the cell plane during cycling of $\psi(t$, as in the thinner cell (see **Fig. 2J**). Scale bars = 200μm.

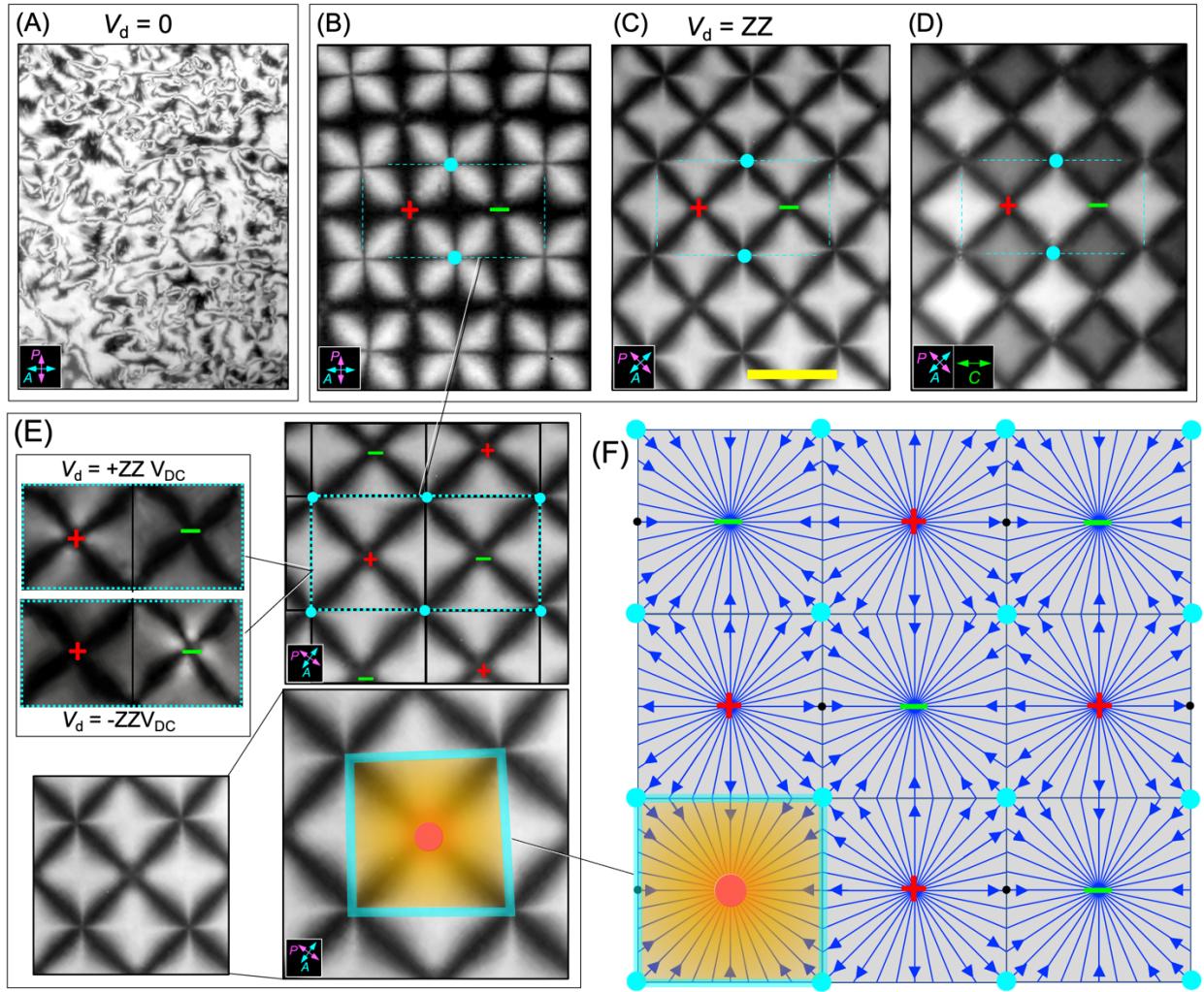

*Figure 5*: (*A*) Initial random planar texture in a bare ITO, $d = 0.8\mu$m RM734 cell at $T = 100$ °C. (*B-F*) Various images and representations of the FFS texture for a $f = 900$ Hz, $V_p = 5$V triangle wave. The frame rate is 60 Hz so that each of the images is the time average of instantaneous images over ~15 cycles. This texture is two dimensional, with no evidence for twist in the direction $z$ normal to the image plane. (*B*) Polarizer and analyzer along DSK lines. (*C*) Polarizer and analyzer at 45° to the DSK lines. (*D*) Compensator establishes the directions of $(\hat{n}, P, v_u)$ shown in (*E*). (*B,E*) The lattice unit cell (dashed box) comprises a +/- defect pair. (*B,E*) High DC voltage breaks the optical symmetry of the $+2\pi$ and $-2\pi$ defects. (*E,F*) Texture comprises a periodic array of plaquettes in 2D, each plaquette a $+2\pi$ $(\hat{n}, P, v_u)$ defect (red core) bounded by DSK domain walls (*cyan* lines). The DSK lines cross to form topologically required $-2\pi$ defects. Scale bar = 300$\mu$m.

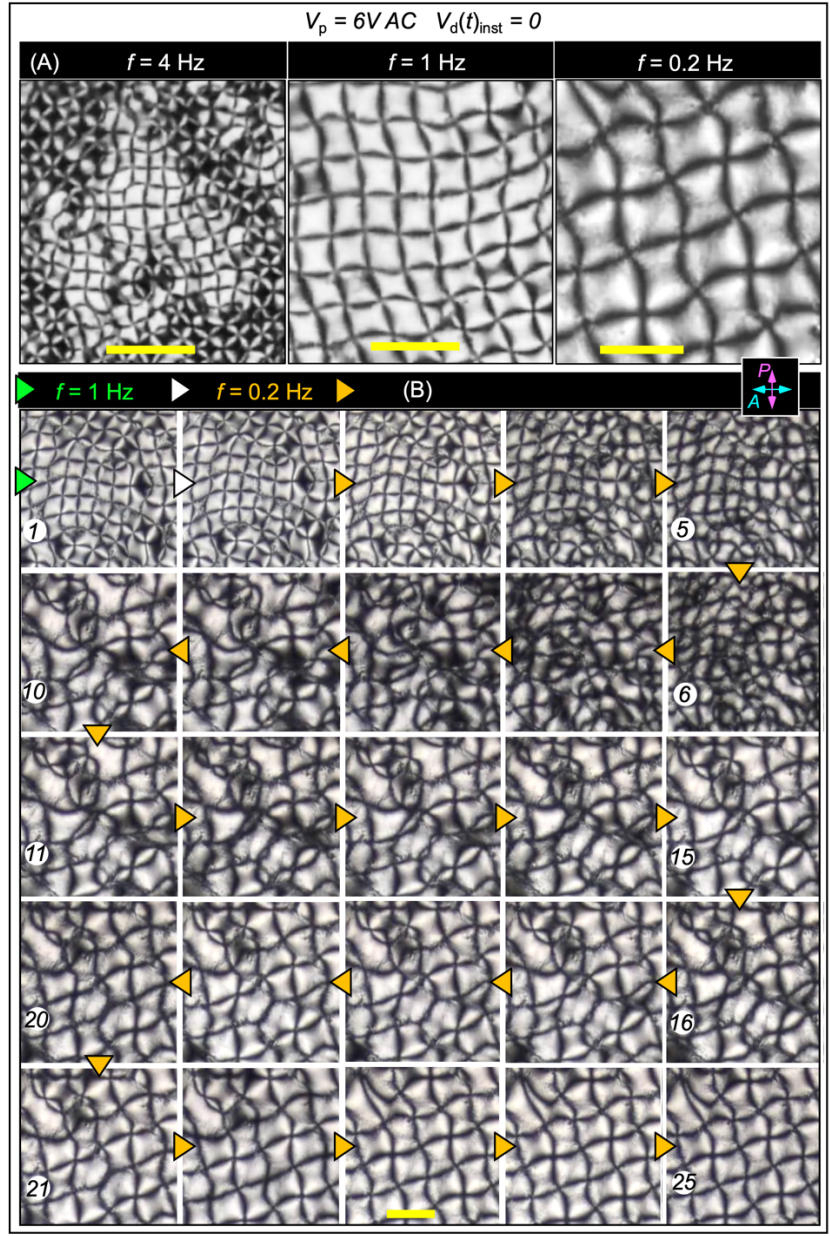

*Figure 6*: FFS textures obtained in a bare ITO, $d = 1.0$ μm RM734 cell at $T = 43$ °C, with a drive voltage amplitude $V_P = 6$V AC and $f = 4$ Hz. The high viscosity of RM734 and the surface memory obtained in these cells at lower temperatures stabilize FFS textures at such low frequencies, as well as enabling video observation at standard frame rates (60 fps) of the orientation of the $\hat{n}$ (*r*,*t*),**P**(*r*,*t*) in response to the instantaneous drive voltage, $V(t)$, as shown here in the sequence of DTOM transmission images, 1-20. This sequence starts with $V(t) = 6$ V giving extinction at $|\psi(t)| = 90°$ (frame 1) and steps with equal time intervals through a single complete cycle of $V(t)$, ending at frame 20. The brightest images are those frames grabbed when $V(t)$, $\psi(t) = 0$. Scale bars = 100μm.

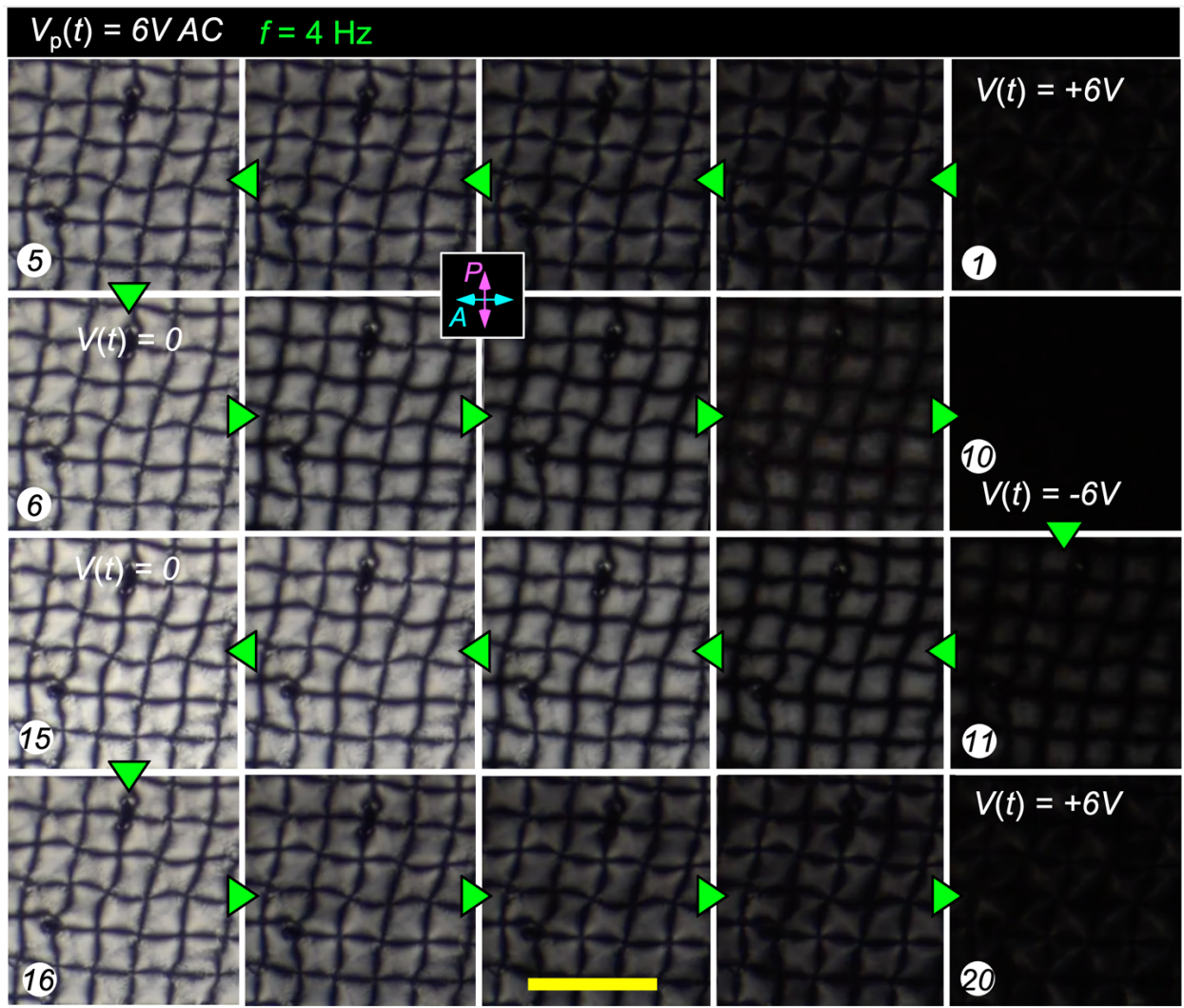

*Figure 7*: FFS textures obtained in a bare ITO, $d = 1.0$ μm RM734 cell at $T = 43$ °C. All of these images are frames grabbed when $V(t), \psi(t) = 0$. They show that FFS lattices can be stabilized at the low frequencies indicated, a result of the high viscosity of RM734 and of the surface memory obtained in these cells at lower temperatures. (**A**) Typical patterns at $f = 4$, 1, and 0.2 Hz, showing a monotonic increase of plaquette size, $L$, with decreasing $f$, approximately as $L \propto (1/f)^{½}$. (**B**) This slow dynamics enables real-time observation of the response of such lattices to change in frequency. Frame 1 shows a stable starting lattice texture at $f = 1$ Hz (green arrow). The frequency is switched to between frames 1 and 2 (white arrow), beginning a transformation ending with a final stable lattice with ~2X the lattice cell size (orange arrows). Starting from frame 2, a subsequent frame is grabbed at each time the instantaneous $V(t), \psi(t)$ again crosses zero, i.e. at intervals of 2.5 sec. Changes in the initial lattice are observable even after a single cycle of the $f = 0.2$ Hz drive. Scale bar = 300μm.

*Figure 8*: FFS textures obtained in a bare ITO, $d = 1.0$ μm RM734 cell at $T = 43$ °C. (**A**) FFS pattern stabilized by $V_P = 6$V, $f = 0.5$ Hz drive, imaged when $\psi(t)$ is passing through $V(t) = 0$, $\psi(t) = 0$.
(**B**) When the voltage is switched to $V = 0$ (blue arrow), $\hat{n}(r,t), P(r,t)$ relaxes over several minutes to a surface controlled texture which is planar but disordered, and which has with an orientational bias toward the pattern of (*A*). (**C**) The voltage is restarted between (*B*) and (*C*) (magenta arrow), passing through $|\psi(t)| = 90°$ (the extinction orientation) in (*C*). (*D*) Pattern is almost completely reestablished after passage through single half-cycle #1, through (*C*) to $\psi(t) = 0°$ in (*D*), where the bright image is also made passing through. Continuing similarly through the next half cycle (#2) makes the DSK dark lines somewhat sharper, perfectly recreating the FFS texture of (*A*). This shows that the flow-induced torques are in control of the FFS, overwhelming those of the surface during each cycle. Scale bar = 300μm.

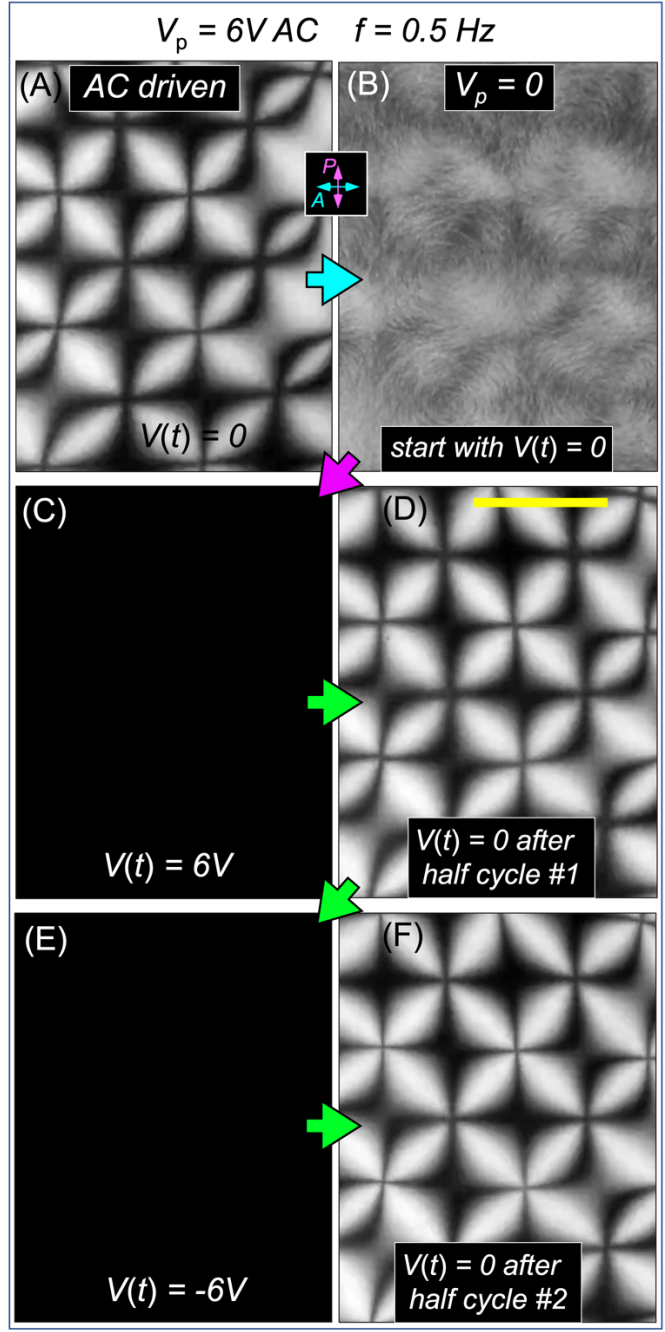

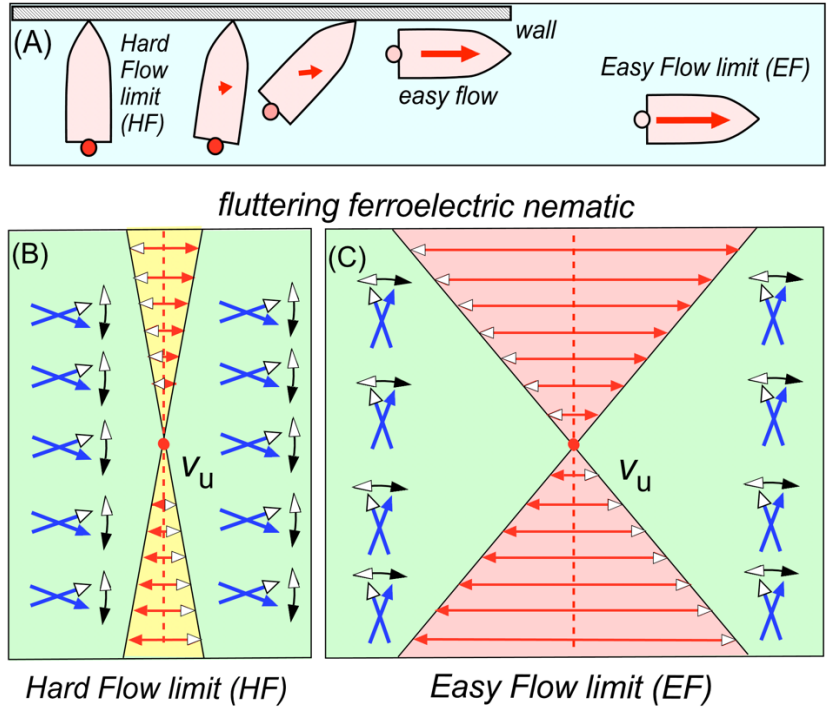

*Figure 9*: (*A*) A floating motorboat moved by an underwater flipper-like paddle. With the boat normal to a smooth wall the motor provides maximum motive power but the boat has no velocity, the "hard-flow (HF)" limit of highest dissipation. If the boat turns a little the system will push itself to the lowest dissipation, "easy-flow (EF)" limit, in this case moving in open water. The nonzero time-average elastic-like turning forces come from the difference in the net reactive force exerted on the boat by the wall between the forward and return strokes of the paddle. (*B,C*) Comparison of the two limiting geometries of uniform $\psi(r)$ fluttering ferroelectric nematic, the resting $\hat{n},P$ being either parallel to the plates [the *planar* (HF) case], or normal [the *homeotropic* (EF) case]. We assume slip boundary conditions for simplicity, the flow allowed by the slab and symmetry being a linear shear with $v_u$ parallel to the surfaces and zero stress $\sigma_{uz} = 0$, as sketched in *Fig. 1E*. (*B*) Each fluttering molecule serves as a paddle which is *least* efficient when driving shear with the velocity field *parallel* to $\hat{n}$, as in the *planar* case, as observed in the FFS textures here. This makes *planar* alignment of $\hat{n},P$ the limit of hardest flow (HF) for uniform $\psi(r)$, with normalized shear gradient $G^{HF} \propto (\alpha_3/\eta_2) \approx 0.28$, and effective driven viscosity $\gamma_{HF} = \gamma_1[1 - (\alpha_3/\eta_2)^2] \sim 0.92\gamma_1$. (*C*) *Homeotropic* alignment of $\hat{n},P$ is the limit of easiest flow (EF) for uniform $\varphi(r)$, with normalized shear gradient $G^{HF} \propto (\alpha_2/\eta_1) \approx 0.80$, and effective driven viscosity $\gamma_{EF} = \gamma_1[1 - (\alpha_2/\eta_1)^2] \sim 0.36\,\gamma_1$. Viscosity ratios are estimates.

*Figure 10*: (*A*) Example of as-grown-in area of a Fluttering Ferroelectric Smooth (FFS) texture in *Fig. 2C* in a $d = 0.8$ μm cell of RM734 at $T = 100$ °C. The principal common features of 2D bulk areas of FFS textures are found here, including: the dominant $+2\pi$ point topological defect core (red) having a radial $\hat{n}$,*P* field of pure splay (orange/yellow), with a nearly complete absence of other defects or irregularities; $N_F$/air boundary interfaces where $\hat{n}$,*P* is normal to the interface line (magenta); distinct linear domain boundaries [dissipation stabilized kinks (DSKs)] where the component of *P* normal to the wall is continuous and 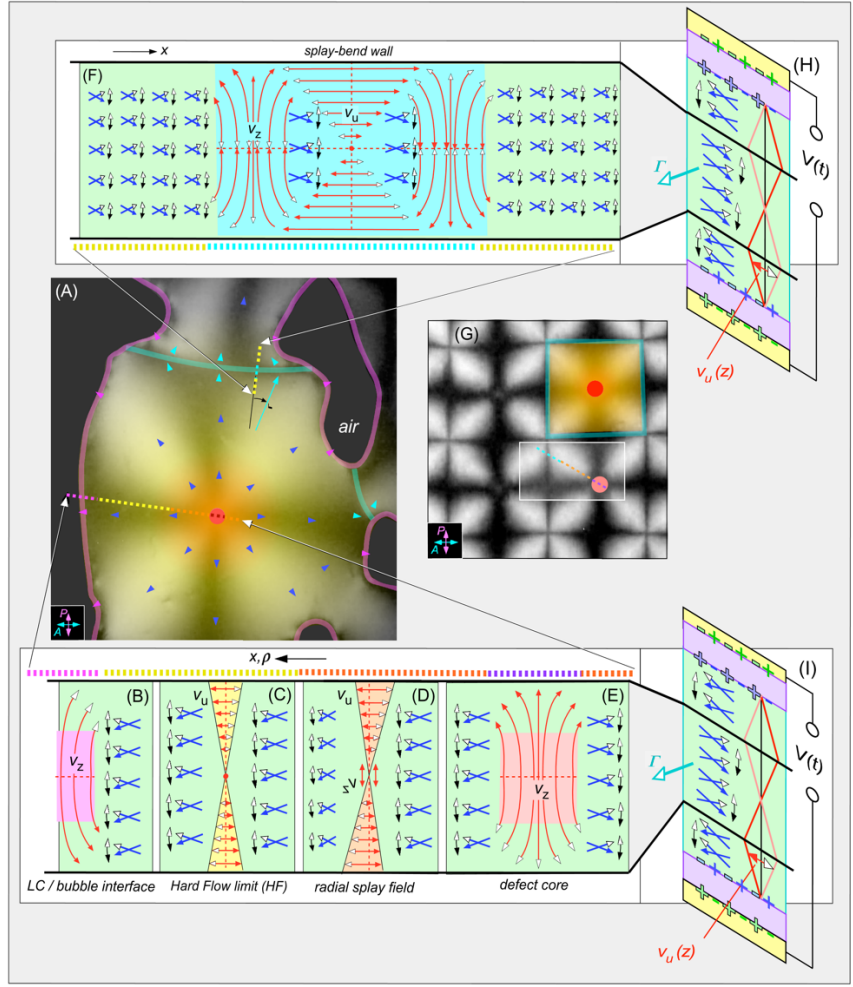 the component parallel to the wall changes sign (cyan). (*B-F*) Section drawings along various trajectories of dashed lines as indicated (*Easy Flow* - magenta, orange, purple, cyan; *Hard Flow* - yellow). The easy flow velocity direction is $v_z$: (*B*) magenta line – The air/NF interface line showing the shaded volume at the boundary where $v_u$ abruptly decreases in which easy flow along $v_z$ drives a ($v_x,v_z$) velocity roll at the interface. The enhanced $v_x$ at the interface orients $\hat{n}$,*P* to be normal to the interface line; (*C*) yellow line – The nearly uniform $\varphi(r)$ region locked into the HF state. Here the easy flow along $v_z$ is cancelled by translational symmetry. (*D*) orange line – The region dominated by splay of $\varphi(r)$, where $\partial v_z(\rho,z)/\partial z = v_\rho/\rho = [v_u = G_{fl}^{HF} d/2]/\rho$. (*E*) purple line – The defect core region showing the shaded volume in which easy flow along $v_z$ drives a ($v_\rho,v_z$) toroidal velocity roll centered on $\rho = 0$. (*F*) cyan line –Dissipation Stabilized Kink (DSK) domain wall between adjacent $+2\pi$ defect areas. Projection of $v_u$ along the line normal to the wall steps discretely up then back down on passing through the wall, creating an easy flow region in the center, forming a ($v_x,v_z$) roll. Pairs of DSK walls cross at 90° to each other to make topologically-required $-2\pi$ disclinations.

Xi Chen[1], Cory Pecinovsky[2], Eva Korblova[3], Matthew A. Glaser[1], Leo Radzihovsky[1]
Joseph E. Maclennan[1], David M. Walba[3], Noel A. Clark[1]*

[1]*Department of Physics and Soft Materials Research Center,*
*University of Colorado, Boulder, CO 80309, USA*

[2]*Polaris Electro-Optics, Inc.*
*3400 Industrial Lane, Suite 7C, Broomfield, CO 80020, USA*

[3]*Department of Chemistry and Soft Materials Research Center,*
*University of Colorado, Boulder, CO 80309, USA*

*Abstract*

Polarization flutter, produced by an applied AC electric field drives an equilibrium ferroelectric nematic ($N_F$) liquid crystal (LC) through a transition into a dissipative active ferroelectric nematic state exhibiting strong elasto-hydrodynamic intermolecular interaction. In such a fluttering ferroelectric, the typical equilibrium $N_F$ textural features adopted to reduce electrostatic energy, such as preferences for director bend, and alignment of polarization parallel to LC/air interfaces, are overcome, giving way to nonequilibrium conjugate structures in which director splay, and alignment of polarization normal to $N_F$/air interfaces are preferred. Viewing the latter textures as those of an active nematic phase reveals that self-organization to reduce effective viscosity and resulting dissipation generates a flow-driven apparent nematic elasticity and interface structuring that dominates equilibrium LC elastic and surface forces.

*Section S1 - Nematic Liquid Crystals (LCs)*

*Dielectric nematics ($N_D$)* - The classic dielectric (non-ferroelectric) nematic liquid crystal (LC) is a bulk fluid material characterized as: (*i*) having long-ranged quadupolar ordering of molecular long axes, describable by a second rank tensor **Q**(*r*) giving the local average molecular long axis and corresponding local optical uniaxis orientation, along the unit vector director $\hat{\mathbf{n}}(r,t)$, with $\hat{\mathbf{n}}$ and -$\hat{\mathbf{n}}$ giving equivalent descriptions of the structure. (*ii*) an elastic medium described by the Frank theory of local elastic spatial variation of its tensor ordering; (*iii*) a hydrodynamic medium in which flow has a macroscopic anisotropic coupling to molecular reorientation and to internal forces applied within the fluid. The dielectric nematic is not macroscopically polar and so the torque applied to **Q**(*r*) by an external electric or magnetic field favors a single molecular configuration, independent of the sign of the field . Such a mechanism was sufficient to create the LC display technology that enabled the portable computing revolution, in which bright and dark pixels were, for example, field-on or field-off states and bright-to-dark switching was achieved by overdamped viscoelastic relaxation of the LC [1].

*Ferroelectric nematics ($N_F$)* - The discovery of liquid crystal ferroelectrics (FLCs) in tilted chiral smectics [2], and visualization of their topological defects and textures [3], initiated the study of fluid ferroelectricity, in which the Goldstone dynamical variable of ferroelectric polarization, $\mathbf{P}(r,t) = P\hat{\mathbf{p}}(r,t)$, is its unit vector orientation field $\hat{\mathbf{p}}(r,t)$ [4]. Of particular importance in this development is the unique nature of the polar coupling of molecular orientation to electric field, with the applied torque/volume, $\mathbf{\Gamma}_E = \mathbf{P}\times\mathbf{E}$, depending on the sign of **E**. With the recent discovery of proper ferroelectricity in nematic LCs [5,6,7,8,9,10], a new realm of fluid ferroelectricity has opened up, in which uniaxially symmetric and spatially homogeneous polar nematic liquids have become available, many with greater than 90% polar ordering of their longitudinal molecular dipoles. The ferroelectric nematic ($N_F$) phase also exhibits a fluid ferroelectric polarization field which has fixed magnitude *P*, and a unit vector orientation field $\hat{\mathbf{p}}$ that varies in space and time, but in the $N_F$, $\hat{\mathbf{p}}$ can be taken as being identical to the unit vector "director" field $\hat{\mathbf{n}}$, the local average of molecular long axis , which is also the optical uniaxis. In the $N_F$ we add to the list of important features of $N_F$ phenomenology: (*v*) linear and very strong coupling of molecular orientation to applied and internal electric fields via $\mathbf{\Gamma}_E$ [8].

## Section S2 - Flow calculation for nematic $\psi$, $G_{fl}$ with, free slip boundary conditions

1) *Miesowicz geometries and Leslie relations [11]:*

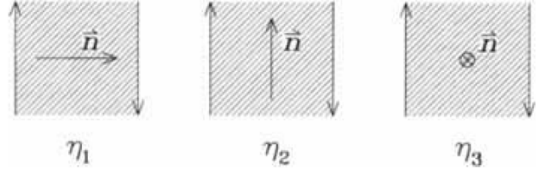

$$\eta_1 = \frac{1}{2}[-\alpha_2 + \alpha_4 + \alpha_5]$$

$$\eta_2 = \frac{1}{2}[\alpha_2 + 2\alpha_3 + \alpha_4 + \alpha_5] = \frac{1}{2}[\alpha_2 + \alpha_4 + \alpha_6]$$

$$\eta_3 = \frac{1}{2}[\alpha_4],$$

where

$$\gamma_1 = \alpha_3 - \alpha_2$$

$$\gamma_2 = \alpha_3 + \alpha_2$$

$$\alpha_2 + \alpha_3 = \alpha_6 - \alpha_5$$

$$\eta_2 - \eta_1 = \gamma_2.$$

2) *Flow alignment angle relative to the **u** axis in **Figure 1** [12,13]:* $\psi_f = \tan^{-1}\sqrt{\frac{\gamma_2+\gamma_1}{\gamma_2-\gamma_1}} = \tan^{-1}\sqrt{\frac{\alpha_3}{\alpha_2}}$.

3) *Coupling of orientation, shear flow, applied torque, and shear stress [14,15]:*

Shear flow equation:

$$\delta(\psi(t))\frac{\partial v_u(z,t)}{\partial z} + \beta(\psi(t))\frac{\partial \psi(t)}{\partial t} = \sigma(t). \quad (S1)$$

Rotation equation:

$$\gamma_1 \frac{\partial \psi(t)}{\partial t} + \beta(\psi(t))\frac{\partial v_u(z,t)}{\partial z} = PE(t)\cos\psi(t) = \Gamma(t) \quad (S2)$$

where

$$\delta(\psi) = \alpha_1 \sin^2\psi\cos^2\psi + \frac{1}{2}[\gamma_2 \cos 2\psi + \alpha_3 + \alpha_4 + \alpha_5] \quad (S3)$$

$$\beta(\psi) = -\frac{1}{2}[\gamma_2 \cos 2\psi + \gamma_1]. \quad (S4)$$

For small amplitude $\psi$ about the average $\langle\psi\rangle = 0$, and defining $G(t) \equiv \frac{\partial v_u(z,t)}{\partial z}$ we have linear equations:

$$\delta(\langle\psi\rangle)G(t) + \beta(\langle\psi\rangle)\dot{\psi}(t) = \sigma(t) \tag{S5}$$

$$\gamma_1\dot{\psi}(t) + \beta(\langle\psi\rangle)G(t) = PE(t). \tag{S6}$$

*Hard Flow (HF) Case* - For the planar aligned fluttering $N_F$ (**Figure 6B**) we have $\langle\psi\rangle = 0$:

$$\delta(0) = \tfrac{1}{2}[\gamma_2 + \alpha_3 + \alpha_4 + \alpha_5] = \eta_2 \tag{S7}$$

$$\beta(0) = -\tfrac{1}{2}[\gamma_2 + \gamma_1] = -\alpha_3. \tag{S8}$$

Then

$$G(t) = [\sigma(t) - \beta(0)\dot{\psi}(t)]/\delta(0) \tag{S9)c}$$

$$\gamma_1\dot{\psi}(t) + \beta(0)G(t) = PE(t). \tag{S10}$$

With a free boundary condition we set $\sigma(t) = 0$, $G(t) = [\beta(0)/\delta(0)]\dot{\psi}(t)$ and:

$$\dot{\psi}(t) = PE(t)[\gamma_1 - \beta(0)^2/\delta(0)]^{-1}. \tag{S11}$$

Typical nematic viscosities are in units of centipoise (cp) are [11,15] $\gamma_1 \sim 76$, $\gamma_2 \sim -78$, $\alpha_3 \sim -1.2$, $\eta_c \sim 103$, in which case $\delta(0) \sim -180$ and $\beta(0) \sim -1$:

$$\dot{\psi}(t) \approx PE(t)/\gamma_1 \tag{S12}$$

$$G(t) = [\beta(0)/\delta(0)]\dot{\psi}(t) = \left[\frac{PE(t)}{\gamma_1}\right]\left[\frac{\alpha_3}{\eta_2}\right] \tag{S13}$$

*Easy Flow (EF) Case* - For the homeotropic aligned fluttering $N_F$ (**Figure 6B**) we have $\langle\psi\rangle = 90°$:

$$\delta(90°) = \tfrac{1}{2}[-\gamma_2 + \alpha_3 + \alpha_4 + \alpha_5] = \eta_1 \tag{S14}$$

$$\beta(90°) = \tfrac{1}{2}[\gamma_2 - \gamma_1] = \alpha_2. \tag{S15}$$

Then:

$$G(t) = [\sigma(t) - \beta(90^o)\dot{\psi}(t)]/\delta(90^o) \tag{S16}$$

$$\gamma_1\dot{\psi}(t) + \beta(90^o)G(t) = PE(t). \tag{S17}$$

With a free boundary condition we have $\sigma(t) = 0$ and:

$$\dot{\psi}(t) = PE(t)[\gamma_1 - \beta(90^o)^2/\delta(90^o)]^{-1} \tag{S18}$$

$$G(t) = [\beta(90)/\delta(90)]\dot{\psi}(t). \tag{S19}$$